\documentclass[aps, pre, superscriptaddress, reprint, onecolumn, 12pt]{revtex4-2}
\usepackage{amsmath, amssymb, graphicx, bm}
\usepackage[colorlinks=true,citecolor=blue,urlcolor=blue]{hyperref}
\usepackage{xcolor, soul}

\begin{document}
\title{Harnessing confinement and driving to tune active particle dynamics}\vspace*{0.5cm}

\author{Aniruddh Murali}
\thanks{Equal contribution}
\affiliation{Simons
Centre for the Study of Living Machines, National Centre for Biological
Sciences, Tata Institute of Fundamental Research, GKVK Campus, Bellary Road, Bengaluru 560065, India}

\author{Pritha Dolai}
\thanks{Equal contribution}
\affiliation{International Centre for Theoretical Sciences, Tata Institute of Fundamental Research, Survey no 151, Shivakote, Hesaraghatta Hobli, Bengaluru 560089, India}

\author{Ashwini Krishna}
\affiliation{Simons
Centre for the Study of Living Machines, National Centre for Biological
Sciences, Tata Institute of Fundamental Research, GKVK Campus, Bellary Road, Bengaluru 560065, India}

\author{K. Vijay Kumar}
\email{vijaykumar@icts.res.in}
\affiliation{International Centre for Theoretical Sciences, Tata Institute of Fundamental Research, Survey no 151, Shivakote, Hesaraghatta Hobli, Bengaluru 560089, India}

\author{Shashi Thutupalli}
\email{shashi@ncbs.res.in}
\affiliation{Simons
Centre for the Study of Living Machines, National Centre for Biological
Sciences, Tata Institute of Fundamental Research, GKVK Campus, Bellary Road, Bengaluru 560065, India}
\affiliation{International Centre for Theoretical Sciences, Tata Institute of Fundamental Research, Survey no 151, Shivakote, Hesaraghatta Hobli, Bengaluru 560089, India}

\begin{abstract}

A distinguishing feature of active particles is the nature of the non-equilibrium noise driving their dynamics. Control of these noise properties is, therefore, of both fundamental and applied interest. We demonstrate emergent tuning of the active noise of a granular self-propelled particle by confining it to a quasi one-dimensional channel. We find that this particle, moving like an active Brownian particle (ABP) in two-dimensions, displays run-and-tumble (RTP) characteristics in confinement. We show that the dynamics of the relative orientation co-ordinate of the particle maps to that of a Brownian particle in a periodic potential subject to a constant force, in analogy to the dynamics of a molecular motor. This mapping captures the essential statistical characteristics of the one-dimensional RTP motion. Specifically, our theoretical analysis is in agreement with the empirical distributions of the relative orientation co-ordinate and the run-times (tumble-rates) of the particle. Finally, we explicitly control these emergent run-and-tumble like noise parameters by external driving. Altogether, our work illustrates geometry-induced tuning of the active dynamics of self-propelled units thus suggesting an independent route to harness their internal dynamics.
\end{abstract}

\maketitle



\section{Introduction}

Active matter is a generic class of nonequilibrium systems wherein the driving and dissipation occur at the level of the constituent individual units. Some of the earliest studies of active matter focused on the hydrodynamic description of collections of self-propelled units to explain coherent phenomenon seen in bird flocks, wildebeest herds, fish schools or insect swarms~\cite{TonerTu1995,TonerTu1998,Toner2005,Ramaswamy2010,MarchettiRMP2013}. The orientational degrees of freedom of the critters are an integral part of their dynamics in these studies, with the interaction between motility and the orientational degrees of freedom leading to complex phases with rich behavior~\cite{MarchettiRMP2013}. However, it is not so much the orientational degrees of freedom nor the anisotropic interactions, but rather the broken detailed balance in the dynamics of the individual units that is the crucial aspect of an active system~\cite{ramaswamyActiveMatter2017}. Minimal models of self-propelled particles that incorporate this essential aspect, in addition to isotropic interactions between individuals, have been of interest in recent years~\cite{BechingerRMP2016}. Synthetic systems, such as self-propelled droplets~\cite{ThutupalliNJP2011,Izri2014}, colloidal systems~\cite{HowsePRL2007,Palacci2013} and granular media~\cite{KumarPRL2011,WalshSM2017}, provide experimental realisations of such \emph{scalar active matter}.

Theoretical models of scalar active matter are typically of three broad types: Active Brownian particles (ABP), exemplified by the motion of self-phoretic colloids~\cite{HowsePRL2007,Palacci2013} and driven granular particles~\cite{KumarPRL2011,WalshSM2017}, move with a constant speed and their motility direction undergoes rotational diffusion. Run-and-tumble-particles (RTP), inspired by the swimming motion of bacteria~\cite{BergNature1972}, also move with a constant speed along a given direction for a certain ``run-time'', after which they undergo an instantaneous ``tumble'' that randomises the direction of motion~\cite{CatesTailleurEPJE2013}. Active Ornstein-Uhlenbeck particles (AOUP) are driven by a self-propelled noise which has a finite-correlation time with a Gaussian strength \cite{fodorHowFarEquilibrium2016}. The essential distinguishing feature in these three models is the characteristics of the active noise driving the self-propelled motion. The changes in propulsion velocity are governed by rotational diffusion for ABPs, by a Poisson process for RTPs, and by an exponentially correlated Gaussian noise for AOUPs. In spite of these contrasting active noise contributions, collections of interacting particles of all three kinds show motility-induced phase-separation (MIPS)~\cite{CatesMIPS2015} and scaling behavior in a single-file geometry~\cite{DolaiSM2020}, pointing to underlying universal features.

In this work, we seek to control the noise characteristics of a scalar active particle and thereby its self-propelled motion. Using an experimental system comprised of a driven granular particle, we demonstrate that laterally confining an ABP-like particle in a narrow channel leads to the emergence of an active noise reminiscent of an RTP-like motion, and that the characteristics of this noise can be controlled by our empirical parameters. To understand this dynamics, the effects of the confining channel are modeled using a simple potential for the orientation of the active particle and we then show that this maps the dynamics of the relative internal coordinate to that of a Brownian particle moving in a periodic potential subject to a constant force. Our results are a clear experimental demonstration of tuning the noise characteristics of an active particle by employing lateral confinement, suggesting new ways to control active particle dynamics.

\begin{figure*}[h!]
\centering
\includegraphics[width=0.75\textwidth]{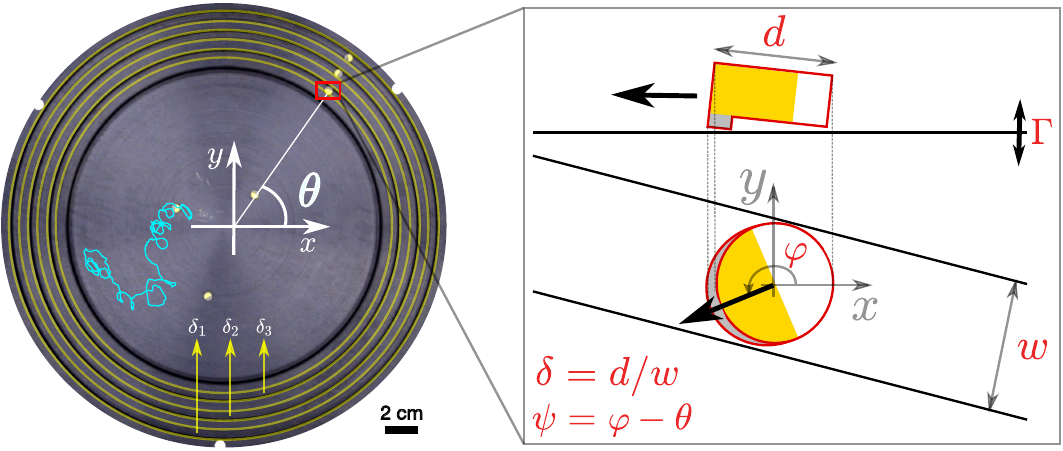}
\caption{Experimental setup: A circular 3-D printed granular particle of diameter $d$ is driven using an electromagnetic shaker (non-dimensional acceleration $\Gamma$). The particle displays active self-propelled motion (blue trajectory) due to frictional asymmetry with the substrate. This two-dimensional motion is well captured by an ABP model describing a chiral particle with position $\mathbf{r}$ and orientation $\varphi$. Quasi one-dimensional confinement of the particle is achieved using concentric circular channels of varying widths $w$ (confinement $\delta = d/w$). This one-dimensional motion is tracked using the co-ordinate $\theta$ and the relative internal co-ordinate $\psi$.}
\label{fig1}
\end{figure*}

\section{Experimental setup}

Our experimental system is comprised of circular 3-D printed disks (particles) of diameter $d$, with an asymmetric leg on one side (Fig.~\ref{fig1} and Supplementary Information); when placed on a flat surface (aluminum disk) and driven using an electromagnetic shaker (Supplementary Information), the particles propel in-plane along a fixed body axis. Owing to its design, a slight shape anisotropy is evident in the 2-dimensional projected diameter of the particle along the propulsion direction (Fig.~\ref{fig1}). In unconfined two-dimensional space, the particle is an active, motile object executing a self-propelled random walk (Supplementary Movie 1) with continuously and noisily varying position $\mathbf{r} = (x,y)$ and orientation $\varphi$ (Fig.~\ref{fig1}, Supplementary Information). In addition to self-propulsion, the particles exhibit chiral motion characterized by an angular speed $\omega$. Altogether, the motion of the active particle in the unconfined 2-dimensional space is well characterized by an ABP model~\cite{WalshSM2017} for a chiral active particle (see MSD analysis in the Supplementary Information): the particle position evolves according to $\dot{\mathbf{r}} = v_0 \; \hat{\mathbf{e}}(\varphi) + \sqrt{2 D_t} \; \boldsymbol{\eta}(t)$ where $\hat{\mathbf{e}}(\varphi) = \cos\varphi \; \hat{\mathbf{x}} + \sin\varphi \; \hat{\mathbf{y}}$ is the local direction of motion, the over-dot indicates a time-derivative, $v_0$ is an active speed, $D_t$ is the translational diffusion coefficient, and the stochastic term $\boldsymbol{\eta}(t)$ is a Gaussian white-noise process with zero-mean and unit-variance. The angular coordinate $\varphi$ of the ABP performs a random walk governed by $\dot{\varphi} = \omega + \sqrt{2D_r} \; \zeta(t)$ with $\omega$ the (chiral) angular velocity, $D_r$ the rotational diffusion coefficient of the ABP while $\zeta(t)$ is a zero-mean and unit-variance Gaussian white-noise process. In our experimental set-up, we are able to control these various parameters of the ABP model, \emph{viz.} $\omega$, $D_r$, $v_0$ and $D_t$, using the driving $\Gamma$ of the electro-magnetic shaker (Supplementary Information).

\section{Emergent stochastic switching dynamics in confinement}

\begin{figure*}[h!]
\centering
\includegraphics[width=\textwidth]{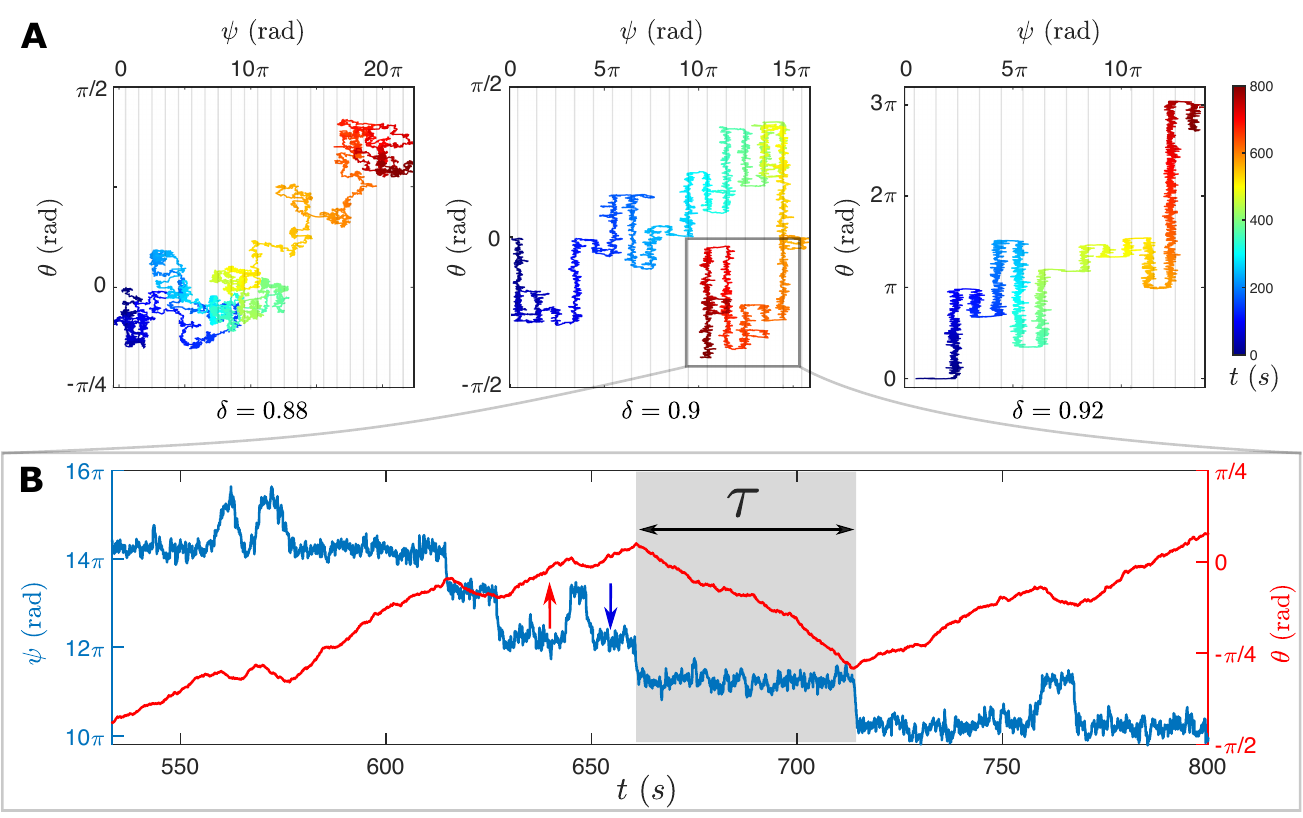}
\caption{Emergent run-and-tumble dynamics in quasi one-dimensional confinement. (A) Parametric plot of the particle trajectories in the translational co-ordinate $\theta$ and the relative orientation co-ordinate $\psi$ for varying confinements $\delta$, with vertical grid lines marking multiples of $\pi/2$. Discrete jumps in $\psi$ are seen at sufficiently high $\delta$.  A constant drift of $\psi$, due to the particle chirality, to the right (``downhill'' direction) is evident. (B) The particle trajectory in $\theta$ is reminiscent of that of a run-and-tumble particle motion with a run-duration $\tau$. Red and blue arrows on the trajectory mark particle flips in the ``uphill'' and ``downhill'' directions respectively.}
\label{fig2}
\end{figure*}

We now turn our attention to the behaviour of these active particles when they are confined to quasi one-dimensional channels -- grooves of width $w$ that run along the periphery of the aluminum disk as concentric circles (Fig.~\ref{fig1}). The geometric configuration of the channels is such that the particles are constrained radially to within $\leq 1\% $ of the channel radius (Supplementary Information) and can still explore all possible values of $\varphi$. When the confinement $\delta=d/w \approx \rm 0.88$ is sufficiently small (\emph{i.e.}, a wide channel) the orientation co-ordinate $\psi$ samples all directions uniformly ($P(\psi)$ for various confinements $\delta$ are not shown) and since this co-ordinate is coupled to the position co-ordinate $\theta$, the particle executes a one-dimensional persistent random walk along the channel (Fig.~\ref{fig2}\textbf{A}, left panel; Supplementary Movie 2). The chirality of particle is evident in the steady drift of the orientation $\psi$. 

A striking difference is manifest when $\delta$ is increased to $\approx \rm 0.9$: instead of evolving continuously, the orientation of the particle fluctuates noisily along one of the channel directions \emph{i.e.} $\psi = \pm \pi/2$, only to stochastically and abruptly switch direction (Fig.~\ref{fig2}\textbf{A} middle panel). The discrete orientation jumps $|\Delta\psi| \sim \pi$ occur either ``downhill'' in a direction dictated by the chirality of the particle or less frequently ``uphill'' (blue and red arrows, respectively in Fig.~\ref{fig2}\textbf{B}). As the confinement $\delta$ is further increased to $\approx 0.92$, the switches in the particle orientation become much less frequent -- in the particular run shown (Fig.~\ref{fig2}\textbf{A}, right panel), they occur only ``downhill''. When $\psi \sim \pm \pi/2$, the particle executes a ``run'' along the channel for a typical ``run-time'' duration $\tau$ until it stochastically switches to a ``run'' in the opposite direction (Fig.~\ref{fig2}\textbf{B}; Supplementary Movie 3).

\begin{figure*}[h!]
\centering
\includegraphics[width=\textwidth]{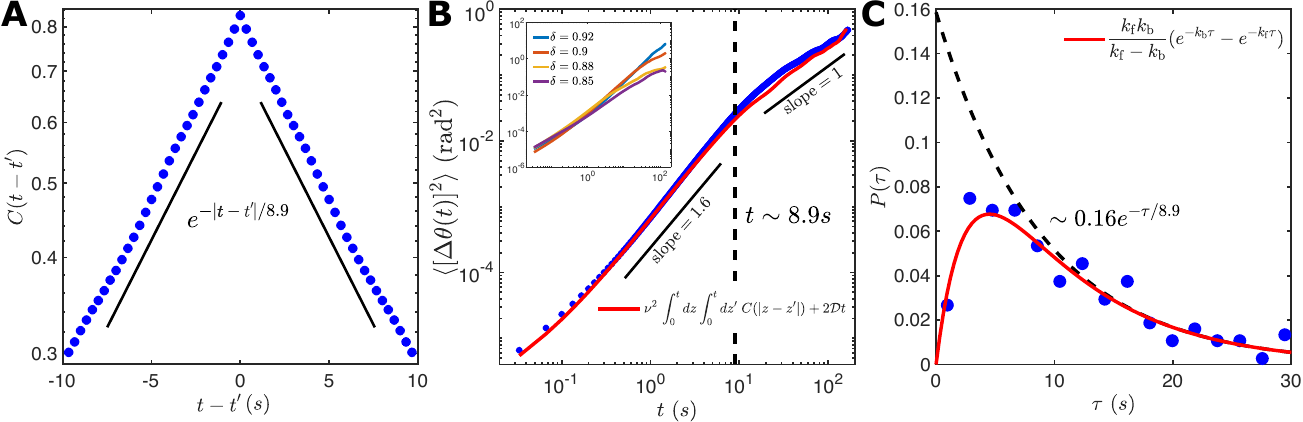}
\caption{Characterisitics of the run-and-tumble dynamics. (A) Two-point correction function  $C(|t-t'|) = \langle \sigma(t) \; \sigma(t') \rangle$ where $\sigma(t) = \sin[\psi(t)]$ is the active noise. (B) The empirical mean-squared-displacement $\langle[\Delta\theta(t)]^2\rangle$ follows the RTP model in equation \eqref{eq:theta_eqn}. MSDs for varying confinements are shown in the inset. (C) The run-time distribution $P(\tau)$ of $\theta$, equivalently the dwell-time distribution of $\psi$, exhibits two-time scales fit by the convolution of the statistics of two Poisson processes (equation \eqref{eq:P_tau}). }
\label{fig3}
\end{figure*}

Taken together, the characteristics discussed above are reminiscent of a run-and-tumble particle in one-dimension, with position co-ordinate $\theta$. As such, the relevant equation describing the dynamics of $\theta$ is that of an ABP confined to move on a circle of radius $R$ (Supplementary Information), \emph{viz.}
\begin{align}
\dot{\theta} = \nu \, \sin\psi + \sqrt{2\mathcal{D}} \; \xi(t) \equiv \nu \, \sigma(t) + \sqrt{2\mathcal{D}} \; \xi(t)
\label{eq:theta_eqn}
\end{align}
where $\nu=v_0/R$, $\mathcal{D}=D_t/R^2$, $\xi(t)$ is a Gaussian white-noise process with zero-mean and unit-variance, and $\sigma(t) = \sin[\psi(t)]$ is the active noise. This model is validited by first computing the empirical two point correlation $C(|t-t'|) = \langle \sigma(t) \; \sigma(t') \rangle$ of the active noise, assuming $\sigma(t)$ is a stationary stochastic process. The correlation function decays exponentially with a time-constant $\tau_{\theta} \sim 9s$ (Fig. \ref{fig3}\textbf{A}). For $\psi \sim \pm \pi/2$, the active noise $\sigma(t) \sim \pm 1$, and thus $\nu$ would correspond to the active speed of the RTP-like motion for $\theta$. The bounded values $\sigma(t) \in [-1,1]$ and the exponential decay of $C(|t-t'|)$ thus justify the effective dynamics of $\theta$ in equation \eqref{eq:theta_eqn} as the position coordinate of an RTP-like particle.

We next compare the empirically measured mean-squared-displacement (MSD) $\langle [\Delta\theta(t)]^2 \rangle$ with that predicted from equation \eqref{eq:theta_eqn}, using the values of $v_0$ and $D_t$ inferred from the two-dimensional experiments for the ABP model:
\begin{align}
\langle [\Delta\theta(t)]^2 \rangle = \nu^2 \, \int_{0}^{t} dz \int_{0}^{t} dz' \; C(|z-z'|) + 2 \mathcal{D} t.
\label{eq:theta_msd}
\end{align}
An excellent agreement is found between the empirical MSD and that given by equation \eqref{eq:theta_msd} (Fig. \ref{fig3}\textbf{B}). Of note in the MSD are the crossovers from an initial diffusive regime (governed by passive translational diffusion), to a super-diffusive regime (governed by the self-propelled motion of the particle), and then, eventually, to a diffusive regime for time-scales $t \gg \tau_{\theta}~(\sim {\rm 9}s)$ corresponding to the decorrelation of the persistent motion driven by the active noise $\sigma(t)$. The effective diffusion coefficient in this asymptotic regime has an active contribution $\sim \nu^2 \, \tau_{\theta}$ in addition to the (angular) translational diffusion coefficient $\mathcal{D}$ \cite{EbbensPRE2010,malakarSteadyStateRelaxation2018}. This additional contribution to the effective diffusivity could dramatically increase with increasing confinement as the particle becomes more RTP-like. Indeed, the effective diffusion coefficient does increase with confinement, and in fact for the highest confinements,  a crossover to the eventual diffusive regime is not seen on the timescale of our experiments (Fig. \ref{fig3}\textbf{B}, inset). These results conclusively suggest that equation \eqref{eq:theta_eqn} is indeed a good descriptor for the dynamics of $\theta$, and also that the parameters describing the motion in confinement remain reasonably similar to the particle motion in two-dimensions (more details in the Supplementary Information).

The empirically measured dwell time distribution of $\psi$ (equivalently, the run time distribution of $\theta$) distribution of $\psi$ is well fit by a bi-exponential function
\begin{align}
P(\tau) = \frac{k_{\rm f} \, k_{\rm b}}{k_{\rm f} - k_{\rm b}} \; \left( e^{-k_{\rm b} \, \tau} - e^{-k_{\rm f} \, \tau}\right).
\label{eq:P_tau}
\end{align}
where $k_{\rm f/b}$ are the forward (``downhill'') and backward (``uphill'') rates of the switching between the states $\psi \sim \pm \pi/2$ (Fig. \ref{fig3}\textbf{C}). It is clear from the above expression that asymptotically, \emph{i.e.,} for $k_{\rm f} \tau \gg 1$ and $k_{\rm f} \gg k_{\rm b}$, the dwell-time distribution $P(\tau) \sim k_{\rm b} \, e^{-k_{\rm b}\tau}$, decaying exponentially with the rate $k_{\rm b}$. For the same driving strength $\Gamma$, the time-scales that govern the decay of the two-point correlation $C(|t-t'|)$ and the crossover of the MSD $\langle [\Delta\theta(t)]^2 \rangle$ from the super-diffusive to the diffusive regime are comparable to the backward (``uphill'') switching rate \emph{i.e.} $1/k_{\rm b} \sim \tau_{\theta}$.

Several remarks are in order. First, it should be noted that unlike the standard RTP model with an exponentially distributed run time between tumbles, our active particle dynamics demonstrates a non-monotonic distribution $P(\tau)$ of run-times $\tau$. Second, the discrete jumps in the trajectory of $\psi$ (Fig. \ref{fig2}\textbf{B}) are reminiscent of the positions of a molecular motor stepping along a polymeric track~\cite{Julicher1997,PurcellPNAS2005}. Third, the fit of equation \eqref{eq:P_tau} (Fig. \ref{fig3}\textbf{C}) suggests the presence of (at least) two characteristic switching rates in the RTP-like dynamics of $\theta$ driven by the active noise $\sigma(t)$. Incidentally, equation \eqref{eq:P_tau} is used to model the dwell time distributions of a two-state molecular motor~\cite{PurcellPNAS2005}. All the above points hint at a possible similarity between the dynamics of our active particle in confinement and those of a processive molecular motor. We next show that this is indeed the case.

\section{Analogy to the stepping dynamics of a molecular motor}

\begin{figure*}[h!]
\centering
\includegraphics[width=\textwidth]{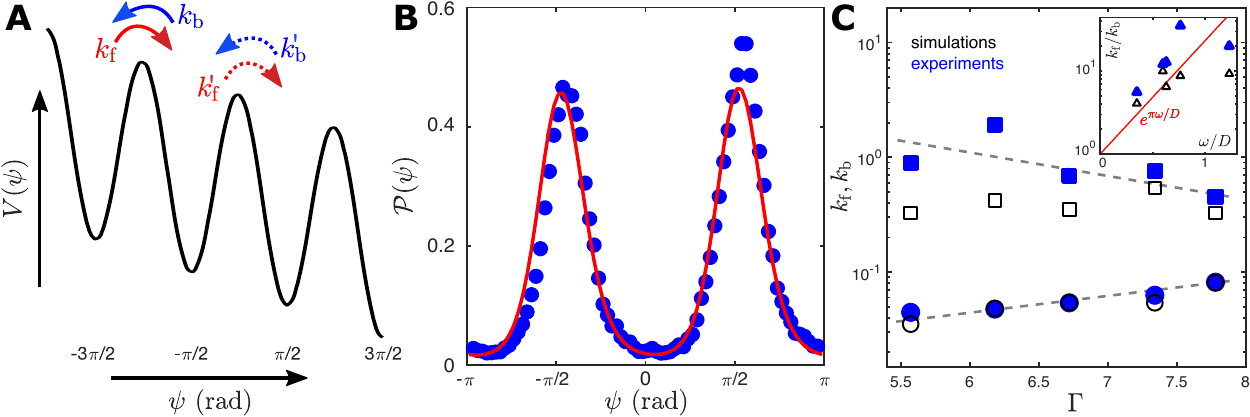}
\caption{Dynamics and control of the stochastic switching. (\textbf{A}) Effective potential for the internal co-ordinate $\psi$ and the transition rates between the neighbouring minima of the potential. In our analysis, $k'_{\rm f/b} \approx k_{\rm f/b}$.  (\textbf{B}) The empirical steady-state distribution $\mathcal{P}(\psi)$ (data points) compared with the analytical solution obtained from the Fokker-Planck equation for the stochastic dynamics of $\psi$ (red line). (C) Control of the forward and backward switching rates, $k_{\rm f}$ and $k_{\rm b}$ by modulating the driving $\Gamma$. Dashed lines are a guide to the eye. Simulations are performed with parameters extracted using a heuristic fitting procedure (SI Text). The inset shows the ratio $k_{\rm f}/k_{\rm b}$ and its correspondence with the prediction from Kramers' rate theory analysis (SI Text).}
\label{fig4}
\end{figure*}

We conjecture that the effects of the confining channel on the particle can be captured by a periodic force of the form $F_{\rm wall} = \lambda \, \sin 2\psi$ in the equation for the orientation co-ordinate, with $\lambda$ being the strength of the wall-particle interaction. In doing this, we simplify the interactions which are quite complicated indeed -- they not only depend on the particle and channel surface roughness and the resultant friction, but possibly also on the amplitude of driving $\Gamma$. However, the step-like trajectories, and the preponderance of the values of $\psi$ close to $\pm \pi/2$, motivate us to test this functional form of $F_{\rm wall}$ to describe the particle motion. Following this intuition, the dynamics of $\psi$ is modelled by the following equation (details in Supplementary Information):
\begin{align}
\dot{\psi} &= \omega + \lambda \, \sin 2\psi - \nu \, \sin\psi + \sqrt{2D} \, ~\overline{\xi}(t)
\nonumber \\
&= -V'(\psi) + \sqrt{2D} \, ~\overline{\xi}(t)
\label{eq:psiDynamics}
\end{align}
where $D=D_t/R^2 + D_r$ with $R$ the (mean) radius of the confining channel, $\overline{\xi}$ is a zero-mean unit-variance Gaussian white noise process, and the prime denotes differentiation with respect the argument of its function. The effective potential
\begin{align}
V(\psi) = -\omega \, \psi + \frac{\lambda}{2} \, \cos2\psi - \nu \, \cos\psi
\label{eq:effective_potential}
\end{align}
thus governs the dynamics of $\psi$. Notice that $U(\psi) = V(\psi) + \omega \, \psi$ is a periodic function of $\psi$. Thus, the Langevin equation \eqref{eq:psiDynamics} describes the stochastic dynamics of a Brownian particle with a position coordinate $\psi$ moving in a periodic potential $U$, subject to a constant ``force'' $\omega$ (Fig. \ref{fig4}\textbf{A}). It is straightforward to solve the Fokker-Planck equation for $\mathcal{P}(\psi)$ in the steady-state, corresponding to this motion (Supplementary Information). This theoretical $\mathcal{P}(\psi)$  agrees well with the empirical distribution, relying on a single fitting parameter $\lambda$ (Fig.~\ref{fig4}\textbf{B}, Supplementary Information). However, $\lambda$ depends on the driving $\Gamma$ (Supplementary Information) which is altogether not surprising given that it encapsulates the effective strength of the complicated interactions between the active particle and the confining channel. 

Having concluded that the effective potential $V(\psi)$ is sufficient to capture the steady-state distribution of the internal coordinate $\psi$, we next show that it can also capture the dynamical properties of $\psi$. The two point correlation function $C(|t-t'|)$ computed from simulations of the Langevin equation \eqref{eq:psiDynamics} compares well with empirical measurements (Supplementary Information). Further, the switching rates, $k_{\rm f}$ and $k_{\rm b}$, as inferred from the fits of equation \eqref{eq:P_tau} to the run-time distributions obtained from both the experimental and simulation data compare well. These rates and other characteristics of the RTP-like motion can be controlled by the driving amplitude $\Gamma$ (Fig.~\ref{fig4}C, Supplementary Information). The transition rates governing the jumps of $\psi$ across successive minima of the potential can be explicitly computed in the Kramers' approximation~\cite{Hanggi1990}. When $\nu\neq0$ there are in-principle four transition rates, $k_{\rm f/b} $ and $k'_{\rm f/b}$, across the potential minima (Fig. \ref{fig4}\textbf{A}); however, in the approximation $\nu \approx 0$ (valid for our experiments, Supplementary Information), only two transition rates $k_{\rm f/b}$ remain. The variation of the ratio $k_{\rm f}/k_{\rm b}$ from the empirical (experimental and simulation) data compares well with the analytical results obtained in the Kramers' approximation (Fig. \ref{fig4}\textbf{C} inset). Further, in analogy with two-state dynamics of molecular motors, the chirality of our active particle makes the forward and reverse transition rates different, concomitant with the bi-exponential distribution of $P(\tau)$ (Supplementary Information). Taken together, the effective potential used to describe the confining effects of the channel is a good predictor for the dynamical properties of $\psi$. 

\section{Discussion and Conclusions}

By confining a self-propelled granular particle to a quasi one-dimensional channel, we have demonstrated emergent active noise properties, qualitatively distinct from those for the two-dimensional unconfined motion of the particle. While the two-dimensional dynamics is well described by an ABP model, RTP-like behavior emerges due to lateral confinement. This dynamics of the translational coordinate $\theta$ is driven by the internal coordinate $\psi$ of the particle, which in turn has similarities to the stepping dynamics of a molecular motor; most notably, this manifests in the non-monotonic nature of the dwell-time distribution of $\psi$ which is identical to that found in two-state processive motors. To complete the picture and capture the dynamics of $\psi$, we reduced the effects of confinement to a simple periodic force, thus mapping it to the dynamics of a Brownian particle moving in a periodic potential subject to a constant external force. Rather surprisingly,  the introduction of this periodic force is sufficient to capture all features of the active particle dynamics.

We emphasize that there is no fundamental reason to expect that the approach taken here to describe the active particle dynamics in confinement -- particularly, one that works for microscopic systems, in or close to equilibrium~\cite{CilibertoPRX2017,McCannNature1999} -- should have successfully described the emergent noisy dynamics of our active particle. Our experimental system is athermal, far from equilibrium, and the various interactions between the granular surfaces are complicated, thus possibly requiring \emph{a priori}, a dynamical description more complex than that given by equations \eqref{eq:psiDynamics} and \eqref{eq:effective_potential} for the emergent statistical features of the particle motion. Therefore, the agreement we find between empirical measurements and theoretical analysis, in particular the Kramers' transition rates is remarkable. 

We conclude with a couple of comments. First, the fundamental character of an active particle is the inseperable coupling between its internal (chemical) co-ordinates and its positional degrees of freedom, while being constantly driven out of equilibrium by an energy throughput~\cite{ramaswamyActiveMatter2017}. The resulting mobility in the parametric space of the internal and positional co-ordinates is non-diagonal implying that a net current along one of these co-ordinates necessarily drives a corresponding current along the other co-ordinate -- our experimental trajectories (Fig.~\ref{fig2}) are an explicit manisfestation of this picture. Second, the RTP-like discrete dynamics that we describe here is not directly imposed, for example, using explicit external means~\cite{KaraniPRL2019,FernandezRodriguezNatComm2020,VizsnyiczaiNatComm2017} but is rather an emergent behavior resulting from confinement. This opens up an avenue to exploit the effects of geometric constraints and non-equilibrium driving to control the noise characteristics of active particles.

\section{Acknowledgements}
We acknowledge support from the Department of Atomic Energy, Government of India, under project no. 12-R\&D-TFR-5.04-0800 and 12-R\&D-TFR-5.10-1100, the Simons Foundation (Grant No. 287975 to S.T.) and the Max Planck Society through a Max-Planck-Partner-Group at NCBS-TIFR (S.T.), the Department of Biotechnology, India, through a Ramalingaswami re-entry fellowship (K.V.K.) and by the Max Planck Society through a Max-Planck-Partner-Group at ICTS-TIFR (K.V.K.). We thank Pawan Nandakishore for help with the initial set-up and experiments. S.T. and K.V.K acknowledge discussions during the KITP 2020 online program on Symmetry, Thermodynamics and Topology in Active Matter.

\bibliography{ABP_RTP_Full}

\newpage

\section*{Supplementary Information}

\setcounter{section}{0}

\section{Experimental details}
\label{sec:ExperimentalDetails}

\subsection{Particle, Shaker and Imaging Setup}
\label{sec:setup}

\subsubsection*{Particles} 
\begin{figure*}[h]
\centering
\includegraphics[width=0.6\textwidth]{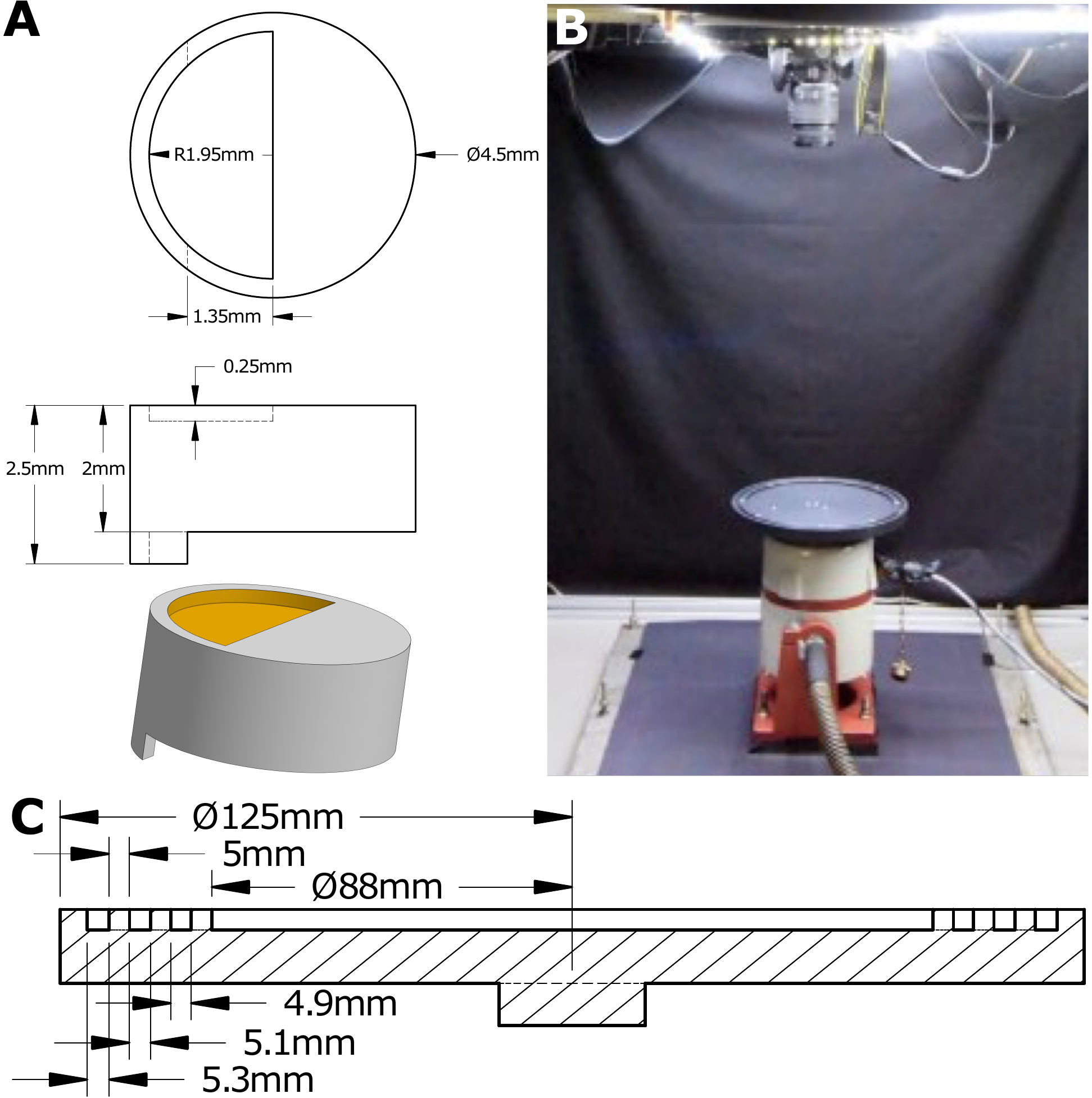}
\caption{\textsf{\textbf{A}} shows the design of our granular self-propelled particle. The overall experimental setup including the electromagnetic shaker and the camera is shown in \textsf{\textbf{B}}. The design of the shaker-head is shown in \textsf{\textbf{C}}.}
\label{fig:particle_shaker}
\end{figure*}

Particles used in the experiment were designed in-house and fabricated using a 3D printer (Form2 SLA 3D printer (minimum resolution of 0.025~mm) from FormLabs which uses FormLab proprietary clear resin for printing). Particles as shown in FIG.~\ref{fig:particle_shaker}~\textbf{A} have a shape asymmetry because of a protruding ``leg'' under the front side of the particle. Particles are 4.5~mm in diameter~($d$) and 2.5~mm in height~($h$) which includes the leg of height 0.5~mm (FIG.~\ref{fig:particle_shaker}~\textbf{A}). The front half of the particle is marked using yellow paint for identification of the particle orientation. 

 \subsubsection*{Shaker setup} 
A Br\"uel \& Kj{\ae}r electromagnetic shaker (LDS V406) is used to excite a patterned aluminum disk on which the particles move. The shaker is suspended on a trunnion mount, which is connected to two sheets of stainless steel plates (FIG.~\ref{fig:particle_shaker}\textbf~{B}). The two stainless steel plates are separated by a rubber pad (hardness 65, shore A)  for isolation of mechanical vibration. Each stainless steel plate weighs 30~kg and the bottom plate is grouted to the ground below. The steel plates along with the mount allow for leveling of the system. The aluminum disk measures 25~cm in diameter and 19~mm in thickness. Circular channels of varying widths of 5.3~mm, 5.1~mm and to 4.9~mm are precision milled into the disk (FIG.~\ref{fig:particle_shaker}~\textbf{C}). The channels are 5~mm in depth with a mean circular radius of 115.7~mm, 105.5~mm and 95.5~mm respectively (from the outer perimeter of the disk inwards, FIG.~\ref{fig:particle_shaker}\textbf~{C}), leaving a two-dimensional arena of radius 88~mm and depth of 5~mm in the center of the disk. A separate head with a single channel of width 5~mm and mean circular radius 116.5 mm was used for some of the experiments reported here. The aluminum disk is polished followed by soft anodization to provide a uniform black background for imaging purposes. The disk is electrically grounded to discharge any static charges present in the system and mounted on the shaker.

\subsubsection*{Imaging} The particle dynamics is recorded using a Nikon DSLR camera aligned above the center of the aluminum disk. The setup is illuminated from above using an array of LED strips (FIG.~\ref{fig:particle_shaker}\textbf~{B}) at the top. The videos are recorded at 30 fps for a duration of 10 -- 20 mins. An opensource program, Digicam Controller, is used to capture videos which are saved directly to a computer.

\subsection{Calibration, Data recording and Image processing}
\label{sec:data_collection}

\subsubsection*{Leveling of the setup} 
\begin{figure*}[h]
\centering
\includegraphics[width=0.8\textwidth]{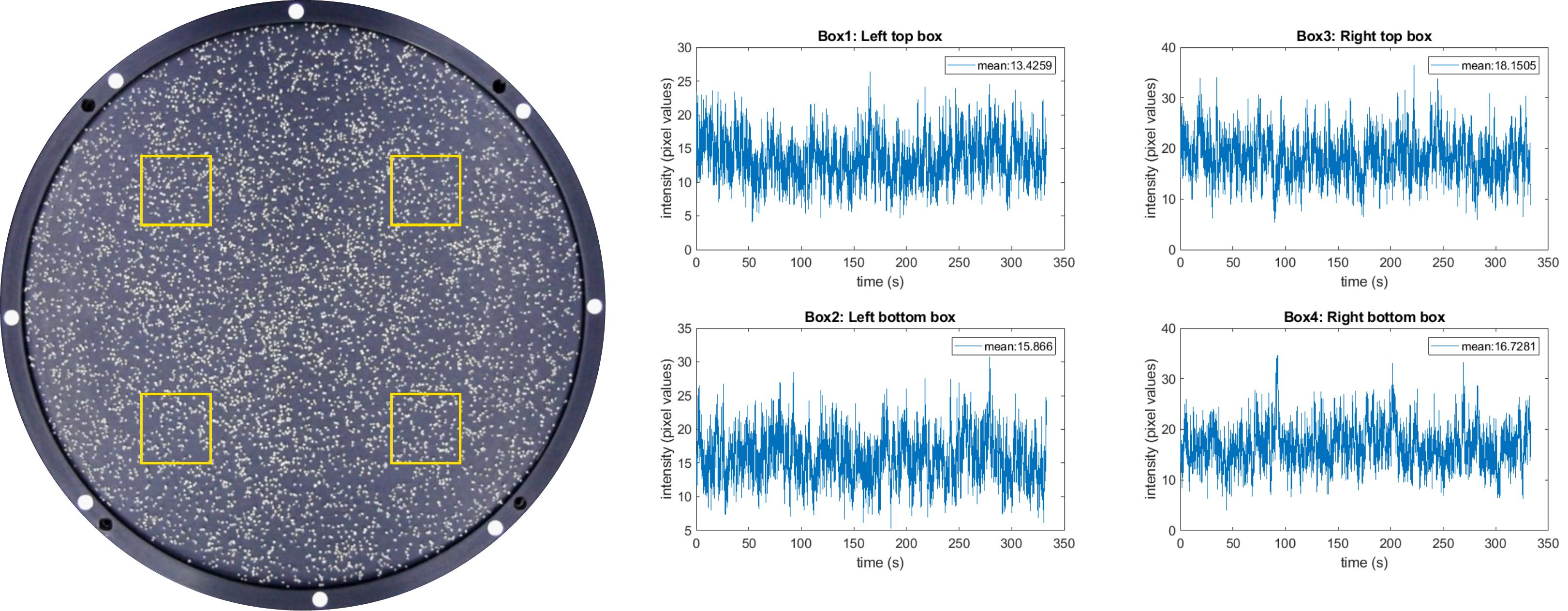}
\caption{To ensure that the head of the shaker is leveled with respect to gravity, we monitor the intensity profiles of fine granular particles (semolina) in four regions of the domain. The intensity versus time plots show that the head is leveled to a good extent.}
\label{fig:level}
\end{figure*}

Granular experiments are sensitive to tilt in any direction and the horizontal level of the aluminum disk needs to be maintained precisely. We use a trunnion mount along one of the axis and screws placed at the corners of the steel plates at the bottom of the electromagnetic shaker to further fine-tune the leveling. To measure the accuracy of the leveling, fine semolina particles are placed in the two-dimensional arena (FIG.~\ref{fig:level}) and excited by the shaker, causing them to disperse in the two-dimensional plane. From a 10 mins video of the dynamics, the image intensity due to the semolina particles is computed in four quadrants of the plate and the pixel intensity difference is used as a measure of the uniformity of the distribution of the particles in the two-dimensional plane which is indicative of the horizontal leveling of the system.

\subsubsection*{Measuring relative acceleration $\Gamma$} To calculate the relative acceleration of the aluminum disk resulting from the actuation, we measure the vertical displacement of the disk using a high speed video recorded at 300 fps (FastCam Mini AX-200, Photron). The high speed camera is positioned perpendicular to the direction of displacements of aluminum disk parallel to the ground and points are marked along the sides of the aluminum disk. The displacement of the points is tracked using image processing. The relative acceleration $\Gamma$  is then calculated using $\Gamma = A\Omega^2/g$, where $A$ is the maximum vertical displacement of the disk, $\Omega$ is the shaking frequency and $g = 9.81 m/s^2$ is the acceleration due to gravity. 

\subsubsection*{Image Processing} 
\begin{figure*}[h]
\centering
\includegraphics[width=0.8\textwidth]{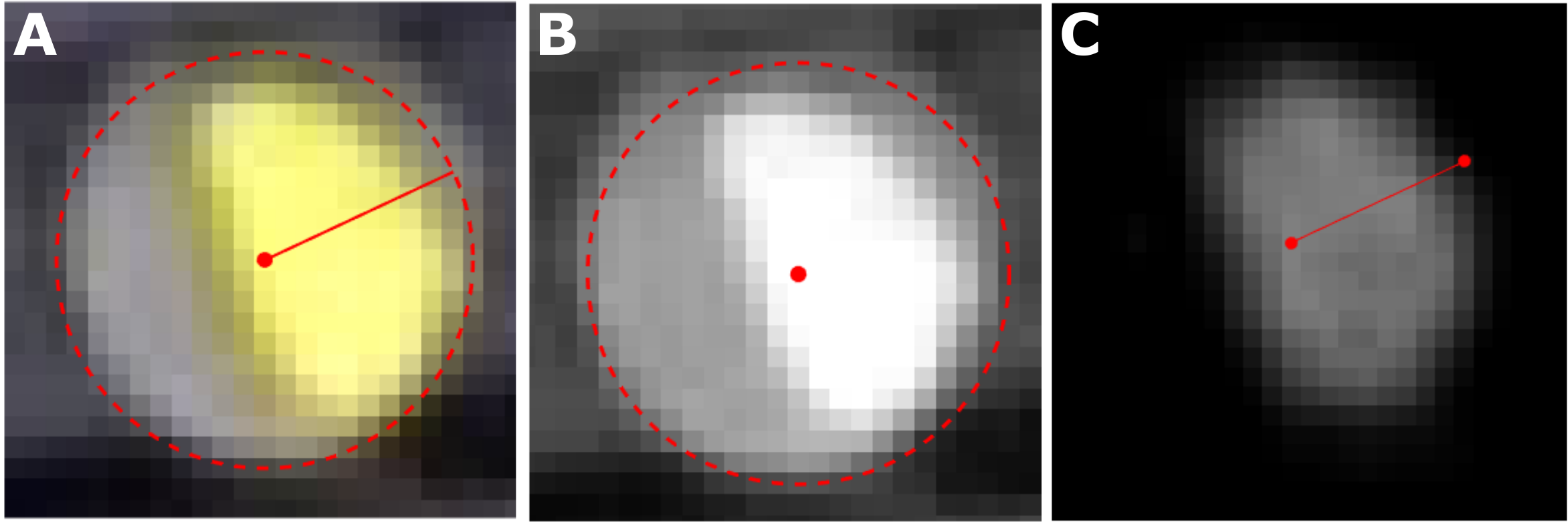}
\caption{The image analysis procedures allow us to detect both the 2D position and the orientation of our active granular particle.}
\label{fig:detectParticle}
\end{figure*}

Videos obtained from the experiments are processed using MATLAB. The video is sliced into frames and a mask is applied to specify the region of interest either to the 1D channel or the 2D arena. Then the image is passed through the \texttt{imfindcircles} function to detect the center of the particle (FIG.~\ref{fig:detectParticle} \textbf{A}). Once the center of particle is detected, tracking is achieved using a least displacement method using custom written MATLAB code. The position of the particle represented by the polar co-ordinate $\theta$ is calculated using the center of disk and the center of the particle and is measured with respect to $x-$axis of the image. To detect the orientation $\varphi$ of the particle, the image is passed through a yellow channel filter to identify the front of the particle with respect to its center (FIG.~\ref{fig:detectParticle} \textbf{B}). The image is then binarized and the \texttt{regionprops} function is used to find the orientation of the semicircular region (yellow mark present at top of the particle, FIG.~\ref{fig:detectParticle} \textbf{C}). Output of the orientation from \texttt{regionprops} is within $[-\pi/2 , \pi/2]$ which are then changed to span $[-\pi , \pi] $. To calculate error in detection, particle positions are recorded without being actuated and their centers and orientations are processed as previously described. From this method, we estimate an error in the detection of $\theta$ to $\pm 0.0017 $ rad and the error in the detection of $\varphi$ to $\pm 0.0436$ rad.


\section{Active Brownian motion in 2D}

We start with the equations for the position $\mathbf{r}$ and the orientation $\varphi$ of an active Brownian particle (ABP) in two-dimensions
\begin{align}
\dot{\mathbf{r}} &= v_0 \, \hat{\mathbf{e}}(\varphi) + \sqrt{2D_t} \, \boldsymbol{\eta}(t),
\label{eq:SI_ABP_r}
\\
\dot{\varphi} &= \omega + \sqrt{2D_r} \, \zeta(t),
\label{eq:SI_ABP_theta}
\end{align}
where the overdot indicates a time-derivative, $v_0$ is the active speed, $\hat{\mathbf{e}}(\varphi)=\cos\varphi \, \hat{\mathbf{x}} + \sin\varphi \, \hat{\mathbf{y}}$ is the instantaneous direction of the active force, $D_t$ the translational diffusion coefficient, $\omega$ the (chiral) angular velocity, and $D_r$ the rotational diffusion coefficient. The stochastic terms $\boldsymbol{\eta}(t)$ and $\zeta(t)$ are uncorrelated Gaussian white noises with zero-mean and unit-variance.

From \eqref{eq:SI_ABP_theta}, it is straightforward to compute the angular mean-squared-displacement
\begin{align}
\langle \left[\Delta\varphi(t)\right]^2 \rangle = \omega^2 \, t^2 + 2 \, D_r \, t,
\label{eq:SI_msd_phi}
\end{align}
where $\langle \cdots \rangle$ indicates an average over the realizations of $\boldsymbol{\eta}(t)$ and $\zeta(t)$. To compute the mean-squared displacement $\langle \left[\Delta\mathbf{r}(t) \right]^2 \rangle$ of the position $\mathbf{r}$, we proceed as follows. The formal solution of \eqref{eq:SI_ABP_r} is
\begin{align}
\Delta\mathbf{r}(t) \equiv \mathbf{r}(t) - \mathbf{r}_0 = v_0 \, \int_0^t ds \;  \hat{\mathbf{e}}[\varphi(s)] + \sqrt{2D_t} \, \int_0^t ds \;  \boldsymbol{\eta}(s)
\end{align}
which leads to
\begin{align}
\langle \left[\Delta\mathbf{r}(t) \right]^2 \rangle &= v_0^2 \int_0^t ds \int_0^t ds' \;  \langle \hat{\mathbf{e}}[\varphi(s)] \cdot \hat{\mathbf{e}}[\varphi(s')] \rangle
+ 2D_t \, \int_0^t ds \int_0^t ds' \;  \langle \boldsymbol{\eta}(s) \cdot \boldsymbol{\eta}(s') \rangle
\nonumber \\
&= v_0^2 \int_0^t ds \int_0^t ds' \;  \langle \hat{\mathbf{e}}[\varphi(s)] \cdot \hat{\mathbf{e}}[\varphi(s')] \rangle
+ 4 \, D_t \, t,
\nonumber \\
&= \frac{2 v_0^2}{\mathcal{D}^2} \left[ \mathcal{D} \, t \; \cos\alpha 
- \cos 2 \alpha  + \cos (\omega t + 2 \alpha) \, e^{-D_r t}\right]
+ 4 \, D_t \, t,
\label{eq:SI_msd_r}
\end{align}
where $\mathcal{D}=\sqrt{\omega^2 + D_r^2}$ and $\tan\alpha=	\omega/D_r$. The intermediate steps for evaluating the double-integral 
\begin{align}
C_{\varphi\varphi}(t) = \int_0^t ds \int_0^t ds' \;  \langle \hat{\mathbf{e}}[\varphi(s)] \cdot \hat{\mathbf{e}}[\varphi(s')] \rangle
\end{align}
are detailed in the next subsection.

\subsection{Evaluating $C_{\varphi\varphi}(t)$}

We simplify
\begin{align}
C_{\varphi\varphi}(t) = \int_0^t ds \int_0^t ds' \;  \langle \cos[\varphi(s) - \varphi(s')] \rangle
\end{align}
To proceed further, we use the formal solution
\begin{align}
\varphi(t) = \omega t + \sqrt{2D_r} \int_0^t ds \, \zeta(s)
\end{align}
of equation \eqref{eq:SI_ABP_theta} and write
\begin{align}
C_{\varphi\varphi}(t) &= \int_0^t ds \int_0^t ds' \;  \left\langle \cos \left[\omega (s-s') + F(s,s') \right] \right\rangle
\nonumber \\
&= \frac{1}{2}\int_0^t ds \int_0^t ds' \; \left[ e^{i \omega (s-s')} \langle e^{i F(s,s')} \rangle + e^{-i \omega (s-s')} \langle e^{-i F(s,s')} \rangle \right]
\label{eq:intermediate_eqn}
\end{align}
where
\begin{align}
F(s,s') = \sqrt{2D_r} \int_{s'}^s dz\, \zeta(z).
\end{align}
We thus need to evaluate $\langle \exp\left[\pm i F(s,s') \right] \rangle$. To do so, we note that $\zeta(t)$ is a zero-mean unit-variance Gaussian white-noise process. This means that $\langle \left[F(s,s') \right]^n\rangle$ vanishes for all odd $n$. And for even $n=2m$, we write
\begin{align}
\langle \left[F(s,s') \right]^{2m} \rangle = (2D_r)^m  \int_{s'}^s \, dz_1 \int_{s'}^s \, dz_2 \ldots \int_{s'}^s \, dz_{2m} \, \langle \zeta(z_1) \zeta(z_2) \ldots \zeta(z_{2m})\rangle.
\end{align}
Furthermore, using the properties of Gaussian stochastic variables to $\langle \zeta(z_1) \zeta(z_2) \ldots \zeta(z_{2m})\rangle$ can be decomposed into $N=(2m-1)(2m-3)\ldots3 \cdot 1$ terms wherein we combine the $z_i$ into all possible pairs. For instance, for $m=2$, we have
\begin{align}
\langle \zeta(z_1) \zeta(z_2) \zeta(z_3) \zeta(z_4)\rangle = \delta(z_1-z_2) \delta(z_3-z_4) + \delta(z_1-z_3) \delta(z_2-z_4) + \delta(z_1-z_4) \delta(z_2-z_3)
\nonumber
\end{align}
and
\begin{align}
\langle \left[F(s,s') \right]^{4} \rangle &= (2D_r)^2 \int_{s'}^s \, dz_1 \int_{s'}^s \, dz_2 \int_{s'}^s \, dz_3 \int_{s'}^s \, dz_4 \; \bigg[ \delta(z_1-z_2) \delta(z_3-z_4)
\nonumber \\
&\qquad
+ \delta(z_1-z_3) \delta(z_2-z_4) + \delta(z_1-z_4) \delta(z_2-z_3) \bigg]
\nonumber \\
&= 3 \, (2D_r)^2 \, (s-s')^2
\end{align}
In general, we get
\begin{align}
\langle \left[F(s,s') \right]^{2m} \rangle = (2D_r)^m  \; (2m-1)(2m-3)\ldots3 \cdot 1 \; (s-s')^m
\end{align}
and thus
\begin{align}
\langle \exp\left[\pm i F(s,s') \right] \rangle &= \sum_{m=0}^{\infty} \frac{(\pm i)^{2m}}{(2m)!} \; \langle \left[F(s,s') \right]^{2m} \rangle
\nonumber \\
&= \sum_{m=0}^{\infty} \frac{(-1)^{m}}{(2m)!} \; (2D_r)^m  \; (2m-1)(2m-3)\ldots3 \cdot 1 \; (s-s')^m
\nonumber \\
&= \sum_{m=0}^{\infty} \frac{(-D_r)^{m}}{m!} \; (s-s')^m
 = e^{-D_r (s-s')}.
\end{align}
Now, we use the above result in equation \eqref{eq:intermediate_eqn} to get
\begin{align}
C_{\varphi\varphi}(t) &= \frac{1}{2}\int_0^t ds \int_0^t ds' \; \left[ e^{(i \omega - D_r)(s-s') } + e^{-(i \omega + D_r) (s-s')} \right]
\nonumber \\
&= \frac{2t}{\mathcal{D}}\cos\alpha -\frac{2}{\mathcal{D}^2}\cos 2\alpha + \frac{2e^{-D_r t}}{\mathcal{D}^2}\cos(\omega t+2\alpha)
\end{align}
where $\mathcal{D}=\sqrt{\omega^2 + D_r^2}$ and $\tan\alpha=	\omega/D_r$ which finally leads to \eqref{eq:SI_msd_r}.

\subsection{Bayesian parameter inference in 2D}
\label{sec:bayesian_2D}

\begin{figure*}[h]
\centering
\includegraphics[width=0.8\textwidth]{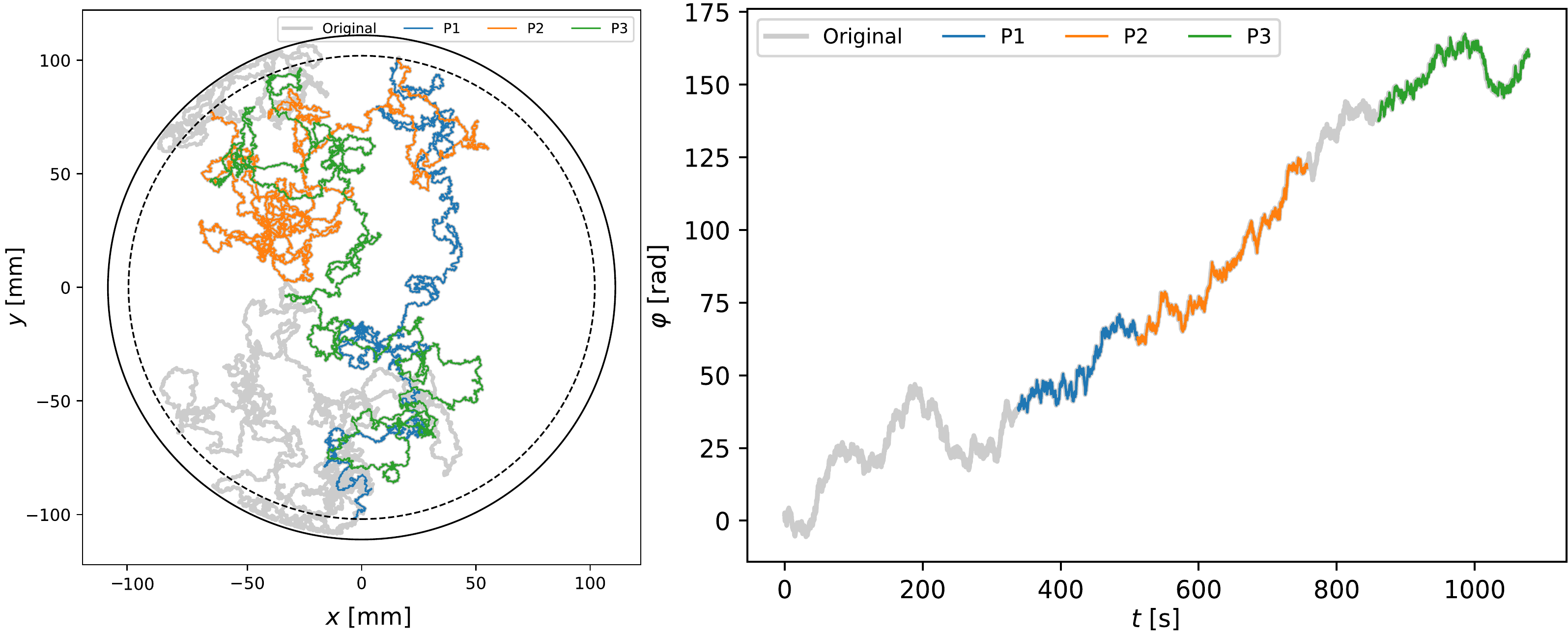}
\caption{We considered particle trajectories in 2D which were at least $150\,s$ and removed the segments which are either shorter than this time or were close to the wall. The gray curves indicates the entire trajectory of a particle while the overlaid color trajectories denote independent segments used in our analysis. }
\label{fig:SI_2D_trajectories}
\end{figure*}

We now compute the mean-squared displacements of the position $\langle \left[\Delta\mathbf{r}(t) \right]^2 \rangle_{\rm expt}$ and orientation $\langle \left[\Delta\varphi(t)\right]^2 \rangle_{\rm expt}$ from experimental data obtained as described in section \ref{sec:data_collection} by considering trajectories that were at least $150 \, s$ long as shown in Fig. \ref{fig:SI_2D_trajectories}. This gave us 25 to 30 statistically independent segments per amplitude $\Gamma$.

\begin{figure*}[h]
\centering
\includegraphics[width=\textwidth]{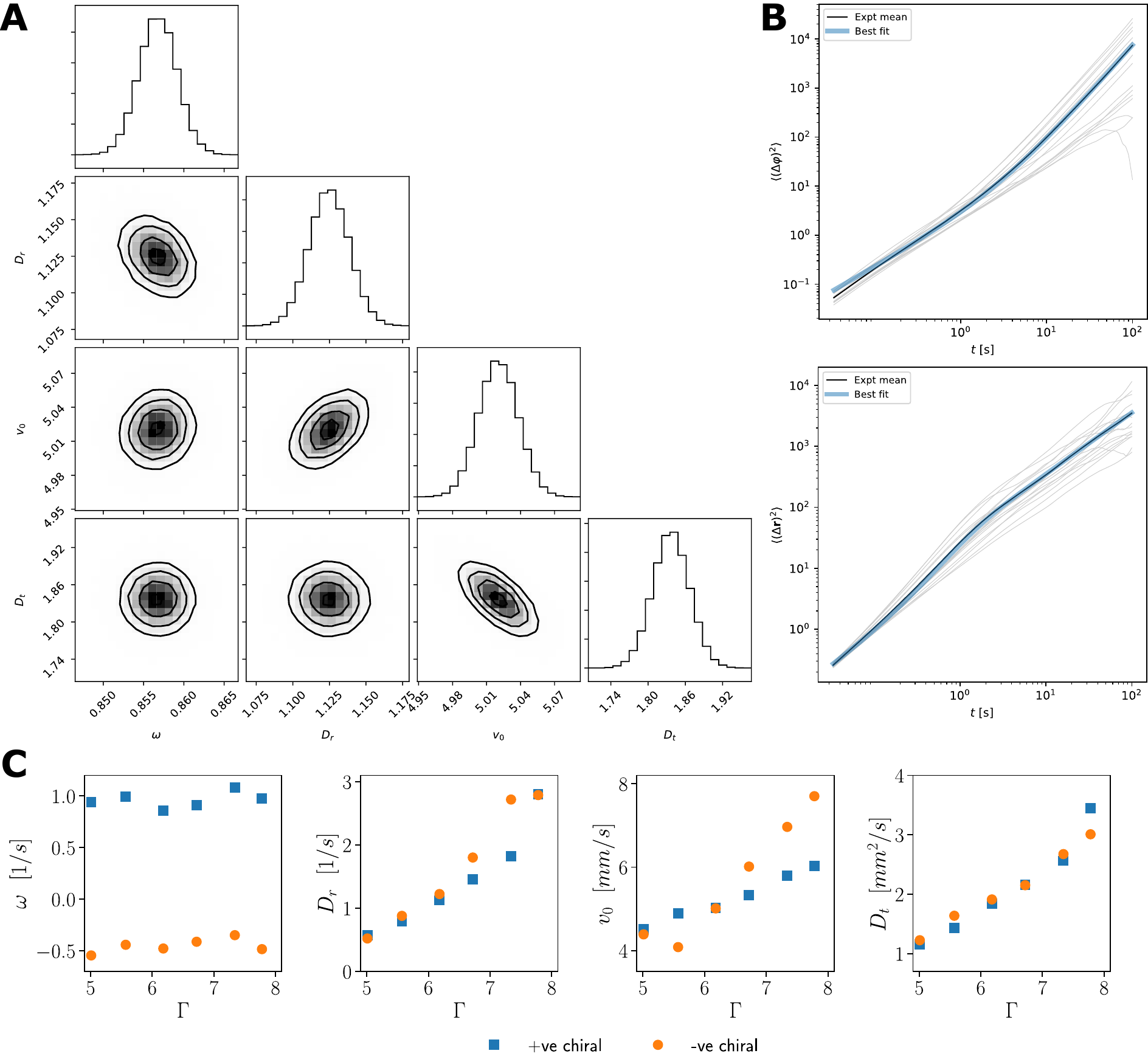}
\caption{\textbf{\textsf{A}} Posterior distributions of the parameters $\omega$, $D_r$, $v_0$ and $D_t$ inferred using a Bayesian analysis of the 2D MSD. The lines indicate confidence intervals of the parameters. In \textbf{\textsf{B}}, we compare the empirical MSD with the theoretical result. The variation of the ABP model parameters with the driving amplitude is shown in \textbf{\textsf{C}}.}
\label{fig:Bayesian_Parameters}
\end{figure*}

Next, we used equations \eqref{eq:SI_msd_phi} and \eqref{eq:SI_msd_r} to infer the parameters of the ABP model that correspond to our self-propelled granular particle. To this end, we performed a Markov-Chain-Monte-Carlo (MCMC) Bayesian inference procedure in the four-dimensional parameter space of $\omega$, $D_r$, $v_0$ and $D_t$. Specifically, we choose flat priors for all the parameters within suitable limits and assumed a Gaussian likelihood function with errors estimated from the independent trajectories mentioned above. We used the Python package \texttt{emcee} \cite{emcee} to implement the MCMC sampling. The posterior distributions obtained in this process are shown in Fig. \ref{fig:Bayesian_Parameters}A while the projection onto the data space of $\langle \left[\Delta\mathbf{r}(t) \right]^2 \rangle$ and $\langle \left[\Delta\varphi(t)\right]^2 \rangle$ is shown in Fig. \ref{fig:Bayesian_Parameters}B. The agreement of the MSD obtained from the parameters inferred using the Bayesian analysis with experimental data clearly shows that the ABP model captures the dynamics of our self-propelled granular particle quite adequately. 

In Fig. \ref{fig:Bayesian_Parameters}C, we plot the dependence of these parameters on the driving amplitude $\Gamma$. As mentioned earlier, our granular particles have a net \emph{chirality}, \emph{i.e.,} $\omega \neq 0$. From Fig.~\ref{fig:Bayesian_Parameters}C, we observe that $\omega$ is nearly constant with the amplitude $\Gamma$. However, the positive and negatively chiral particles in our setup have different values of chirality. The rotational diffusion coefficient $D_r$, the active speed $v_0$ and the translational diffusion coefficient $D_t$ increase with $\Gamma$. 

We have thus mapped the stochastic dynamics of our granular particle moving freely in two-dimensions to an ABP model and have also inferred the model parameters and their dependence on the driving amplitude $\Gamma$. We next turn to characterizing the dynamics of our granular active particle when it is confined to move in a narrow quasi one-dimensional channel.

\section{Active Brownian motion in a confined channel}

\begin{figure*}[h]
\centering
\includegraphics[width=0.4\textwidth]{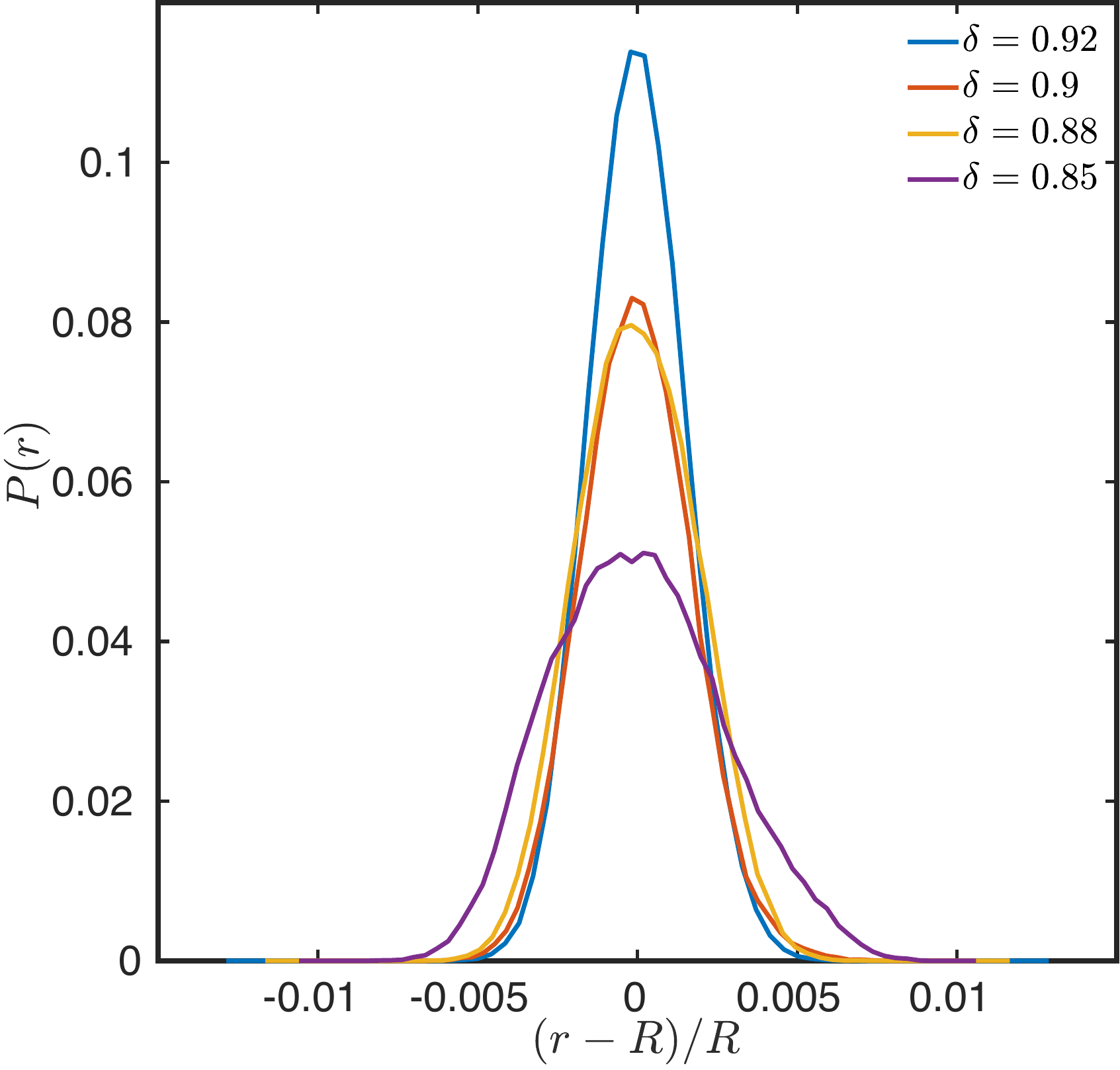}
\caption{Probability distribution of particle displacement in the radial direction for different confinements. We note that the radial motion of the particle is constrained to within $< 1\%$ of the channel radius.}
\label{fig:PrR}
\end{figure*}

To develop a description for the quasi 1D motion in confinement, we first measured the radial motion of the particle in the channel. From FIG.~\ref{fig:PrR}, we see that the motion in the radial direction is constrained to within $< 1\%$ of the channel radius for all the channel widths that we have considered. As such, we can neglect the motion in the radial direction compared to that along the channel.

We transform equations \eqref{eq:SI_ABP_r} and \eqref{eq:SI_ABP_theta} above to plane-polar coordinates $(r,\theta)$ and approximate the motion of our ABP in the confining channel of mean-radius $R$ by assuming that the radial coordinate is a constant at $r=R$. With this approximation, we get an equation for $\theta$ as follows
\begin{align}
R \, \dot{\theta} &= v_0 \, \sin(\varphi-\theta) + \sqrt{2D_t} \; \left[ -\sin\theta \;\; \eta_x(t) + \cos\theta \;\; \eta_y(t) \right].
\end{align}
Since $\boldsymbol{\eta}(t)$ is an uncorrelated Gaussian white noise, we rewrite
\begin{align}
\dot{\theta} &= \nu \, \sin \psi + \sqrt{2\mathcal{D}} \; \xi(t) = \nu \, \sigma(t) + \sqrt{2\mathcal{D}} \; \xi(t),
\label{eq:SI_theta_eqn}
\end{align}
where $\nu=v_0/R$, $\mathcal{D}=D_t/R^2$, the relative orientation coordinate $\psi \equiv \varphi - \theta$, $\xi(t)$ is a Gaussian white noise of zero-mean and unit-variance, and $\sigma(t)=\sin[\psi(t)]$ is the active noise.

As shown in the main text, laterally confining our ABP in a quasi 1D channel leads to the emergence of a bi-modal peak in the probability distribution of $P(\psi)$ of the relative orientation coordinate. As argued in the main text, a simple way to capture the effect of this lateral confinement is to introduce an additional force $\lambda \sin2\psi$ for the angular coordinate $\varphi$. This \emph{confinement-induced force} has a strength $\lambda$ and a periodicity of $\pi$. Thus the equation for $\varphi$ changes to
\begin{align}
\dot{\varphi} &= \omega + \lambda \sin 2\psi + \sqrt{2D_r} \, \zeta(t)
\label{eq:SI_phi_eqn1}
\end{align}
Using \eqref{eq:SI_theta_eqn} and \eqref{eq:SI_phi_eqn1}, we get the following Langevin equation for $\psi$:
\begin{align}
\dot{\psi} = -V'(\psi) + \sqrt{2D} \; \overline{\xi}(t)
\label{eq:SI_psi_eqn}
\end{align}
where $D=D_t/R^2 + D_r$, $\overline{\xi}$ is a Gaussian white noise, and the effective potential is
\begin{align}
V(\psi) = -\omega \, \psi + \frac{\lambda}{2} \cos2\psi - \nu \, \cos\psi.
\label{eq:SI_psi_pot}
\end{align}

\subsection{Brownian motion in a periodic potential}

It is clear from equations \eqref{eq:SI_psi_eqn} and \eqref{eq:SI_psi_pot} that the dynamics of $\psi$ maps onto the dynamics of a  Brownian particle with position coordinate $\psi$ moving in a one-dimensional periodic potential $U(\psi) = \lambda/2 \; \cos2\psi - \nu \, \cos\psi$ with a constant ``force'' $\omega$ applied to it \cite{Julicher1997}. Note, however, that $V(\psi)$ is \emph{not} a periodic function of $\psi$ for $\omega \neq 0$. However, the probability density $\mathcal{P}(\psi,t)$ should be a periodic function of $\psi$ with a period $L=2\pi$.  The Fokker-Planck equation governing the evolution of $\mathcal{P}(\psi,t)$ is
\begin{align}
\partial_t \mathcal{P} = -\partial_{\psi} J,
\label{eq:SI_fpeqn}
\end{align}
where the current
\begin{align}
J = - \left[ \partial_{\psi}V - D \partial_{\psi} \right]\mathcal{P}.
\end{align}
Following \cite{risken1996}, the stationary solution to equation \eqref{eq:SI_fpeqn} is
\begin{align}
\mathcal{P}(\psi) = N \, e^{-V(\psi)/D} 
\left[1 - \frac{I(\psi)}{I(L)} \left(1 - e^{-\omega L/D}\right) \right],
\label{eq:Prob_psi}
\end{align}
with a constant current
\begin{align}
J = \frac{D \, N}{I(L)} \left(1 - e^{-\omega L/D}\right),
\end{align}
and
\begin{align}
\frac{1}{N} = \int_0^{L}d\beta \, e^{-V(\beta)/D} 
\left[1 - \dfrac{I(\beta)}{I(\pi)} \left(1 - e^{-\omega L/D}\right) \right],
\qquad
I(z) = \int_{-L}^{z}dx \,  e^{V(x)/D}.
\end{align}

\subsection{Comparing simulations and experiments}

\begin{figure*}[h]
\centering
\includegraphics[width=0.9\textwidth]{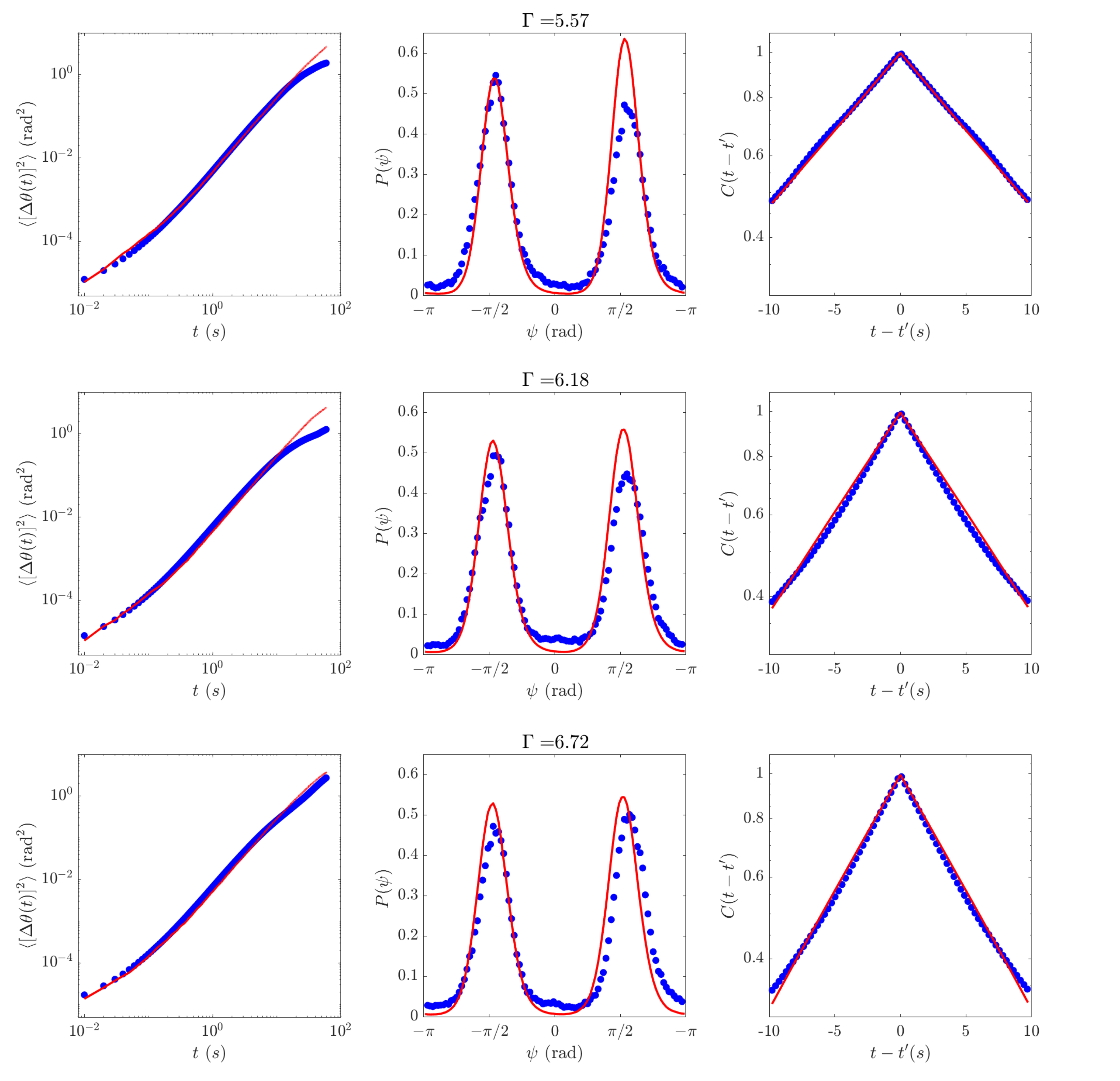}
\caption{Comparing experimental data (filled-circles) for the mean-squared-displacement $\langle [\Delta \theta(t)]^2\rangle$, the steady-state probability distribution $\mathcal{P}(\psi)$ and the correlation function $C(|t-t'|) = \langle \sigma(t) \sigma(t') \rangle$ with Langevin simulation results (lines) of \eqref{eq:SI_theta_eqn} and \eqref{eq:SI_psi_eqn} at various driving amplitudes.}
\label{fig:SI_msd_prob_corr}
\end{figure*}

\begin{figure*}[h]
\centering
\includegraphics[width=0.4\textwidth]{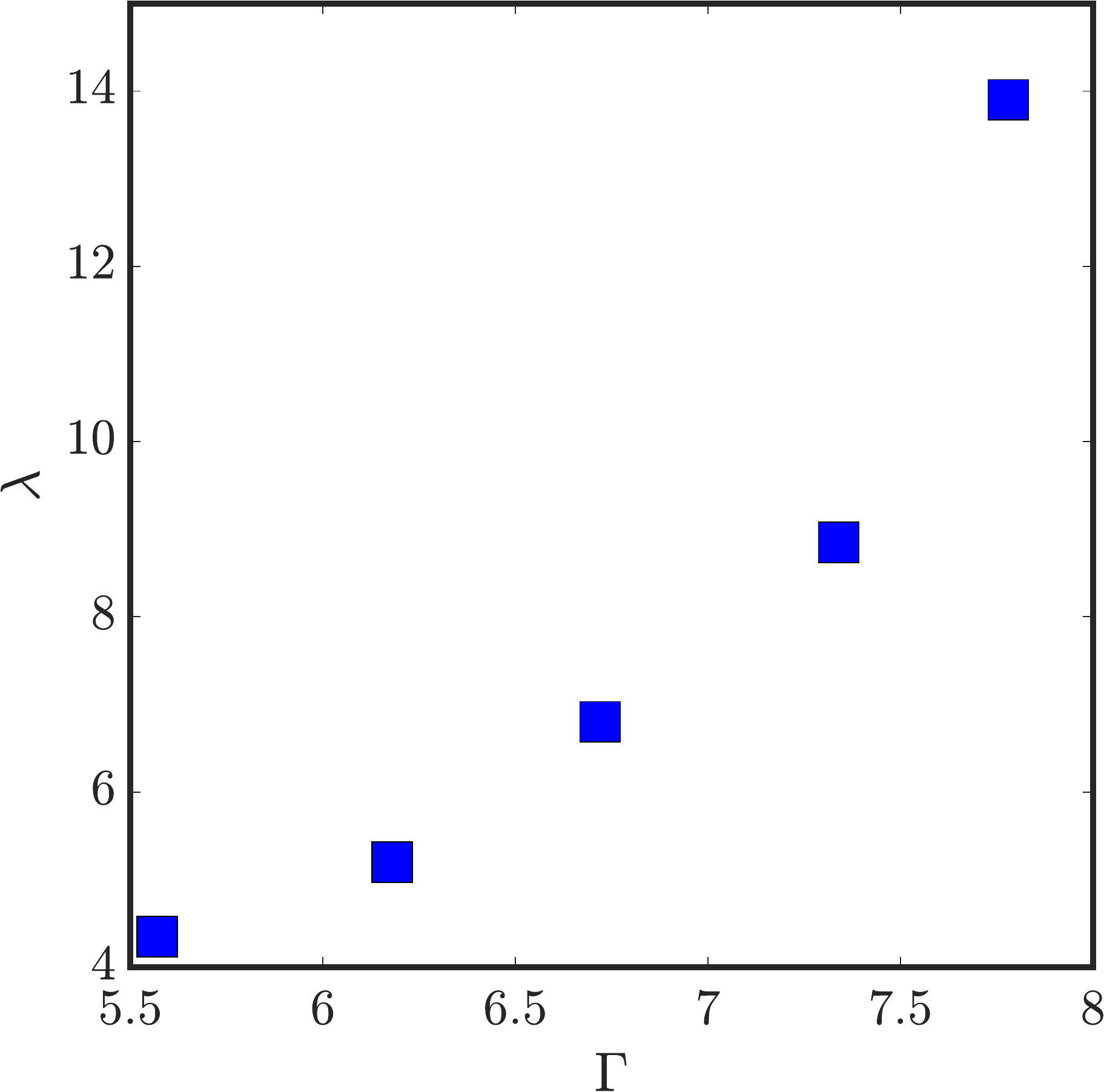}
\caption{Variation of the potential strength parameter $\lambda$ with the driving amplitude $\Gamma$.}
\label{fig:SI_lambda}
\end{figure*}

We performed explicit Langevin simulations of equations \eqref{eq:SI_theta_eqn} and \eqref{eq:SI_psi_eqn} with $\Delta t = 10^{-2} s$ and averaged over $N=10^3$ realizations to get statistical measures of the dynamics. We then asked for the optimal value of $\lambda$ (the potential strength parameter) that would best capture the dynamics of $\langle [\Delta \theta(t)]^2\rangle$, the steady-state probability distribution $P(\psi)$and the correlation function $C(|t-t'|) = \langle \sigma(t) \sigma(t') \rangle$ of the active noise. A heuristic fitting procedure revealed no significant change in the values of $\omega$ and $D_r$ compared to the values inferred from our 2D experiments as reported in section \ref{sec:bayesian_2D}, while $v_0$ and $D_t$ changed by a factor close to 2-3. This is reasonable given our simplistic modeling of the particle-wall interactions in the quasi one-dimensional channel. FIG.~\ref{fig:SI_msd_prob_corr} shows the results of the Langevin simulations compared with empirical results at various driving amplitudes and FIG.~\ref{fig:SI_lambda} shows the variation of $\lambda$ with the driving parameter $\Gamma$.

\section{Flipping Dynamics}

With the agreement between experiments and simulations discussed above, we now compute the transition rates $k_{\rm f}$ and $k_{\rm b}$ from the time-series of $\psi$.

\subsection{Experimental data analysis}

\begin{figure*}[h]
\centering
\includegraphics[width=\textwidth]{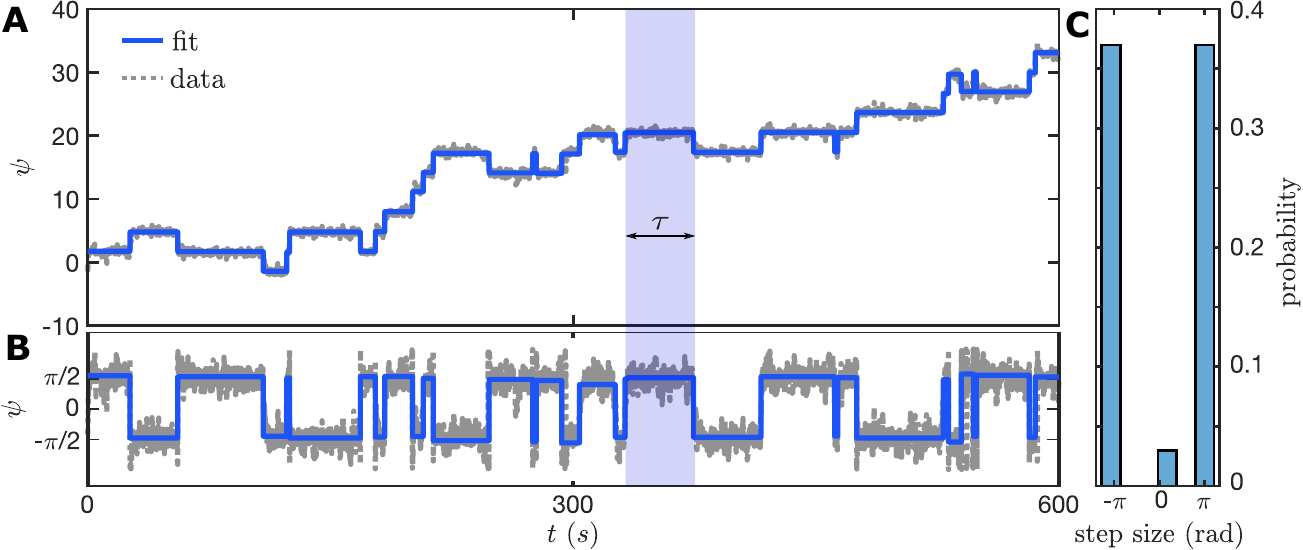}
\caption{Detection of flips in the particle relative orientation $\psi$. \textbf{A}. From the image processing, a time series of the orientation $\psi$ is obtained. This time series is fit using custom MATLAB code to identify discrete jumps/switches in the orientation. The same data is plotted in \textbf{B} where $\psi \in [-\pi, \pi]$. Discrete steps and the preponderance of $\psi \sim \pm \pi/2$ are evident. \textbf{C}. The distribution of the switch/flip sizes shows clear peaks at $ \pm \pi$. Flips corresponding to step sizes close to $0$ are treated as spurious and the detection of the run times $\tau$ and further analysis is performed after such filtering.}
\label{fig:flipsDetection}
\end{figure*}

Using a custom written MATLAB code to implement a previously described algorithm~\cite{Aggarwal2012}, we detect steps in the $\psi(t)$ time series data as shown in FIG.~\ref{fig:flipsDetection}. From the fits obtained to the data, we identify flips (switches) in the ``forward'' (direction of the particle chirality) and ``backward'' directions (FIG.~\ref{fig:flipsDetection}). From the detections, the dwell times $\tau$ in each of the $\psi \sim \pm \pi/2$ states are measured for the particle, and the dwell time distributions $P(\tau)$ are constructed.

\subsection{Waiting time distribution for a composite Poisson processes}

\begin{figure*}[h]
\centering
\includegraphics[width=0.8\textwidth]{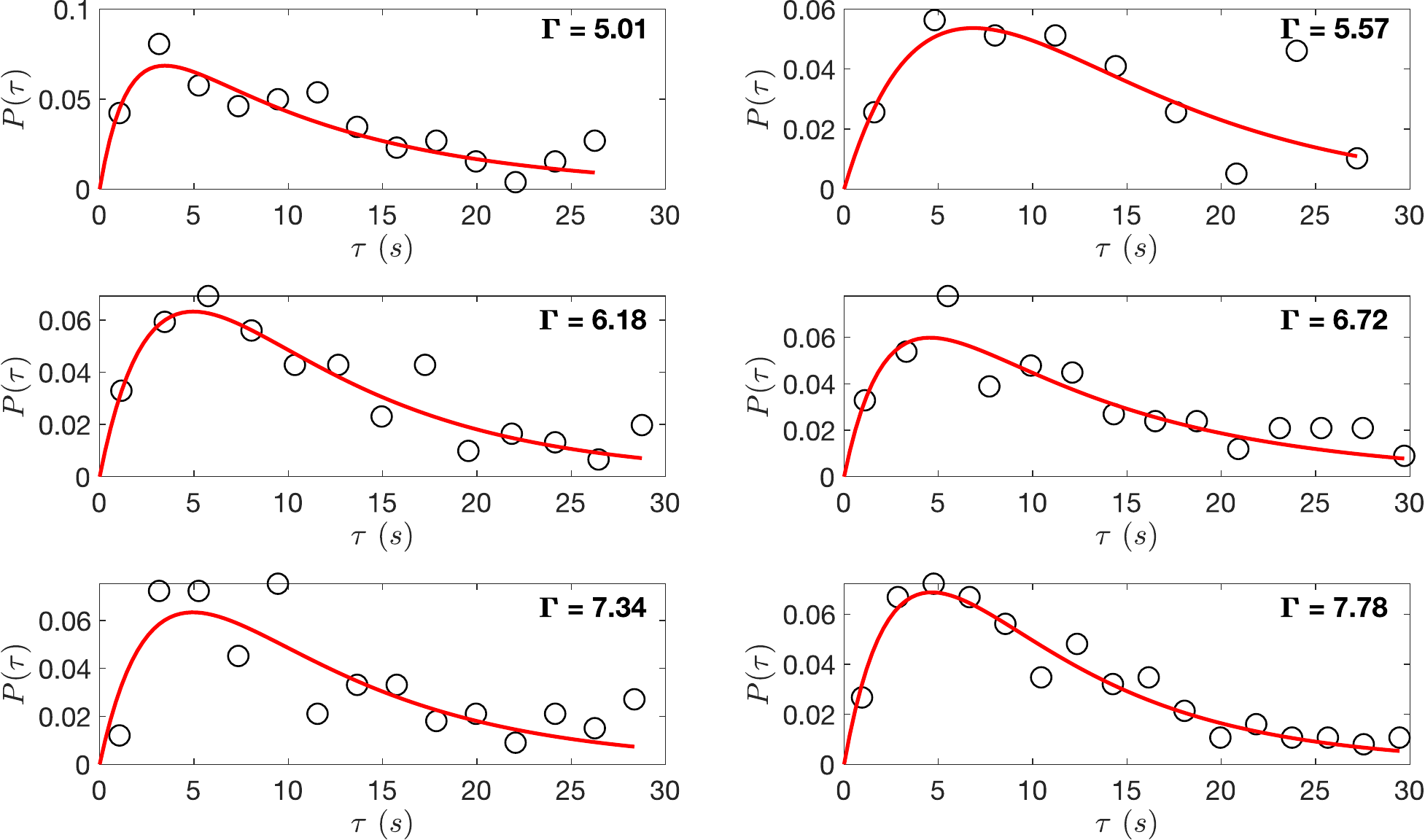}
\caption{The dwell time distribution $P(\tau)$ between the flips seen in the relative orientation coordinate $\psi$. The solid lines are the fit of equation \eqref{eq:waitDist6} to the data. }
\label{fig:ftauDistPanels}
\end{figure*}

The probability density function for the waiting times $\tau$ of a Poisson stochastic process with a rate $k$ is
\begin{equation}
p(\tau) = k \; e^{-k\tau}.
\label{eq:waitDist1}
\end{equation}
Consider two independent Poisson processes, each with distinct rates $k_{\rm f}$ and $k_{\rm b}$. The probability density functions of the waiting times corresponding to each process are
\begin{equation}
p_{\rm f}(\tau) = k_{\rm f} \; e^{-k_{\rm f}\tau},
\label{eq:waitDist3}
\end{equation}
and 
\begin{equation}
p_{\rm b}(\tau) = k_{\rm b} \; e^{-k_{\rm b}\tau}.
\label{eq:waitDist4}
\end{equation}
The waiting time distribution of a stochastic process comprising these independent subprocesses is then obtained via a convolution of \eqref{eq:waitDist3} and \eqref{eq:waitDist4} as
\begin{align}
P(\tau) &= \int_{0}^{\tau}dt \; k_{\rm f} \, k_{\rm b} \, e^{-k_{\rm f} t}e^{-k_{\rm b}(\tau-t)}
= \frac{e^{-k_{\rm b}\tau} - e^{-k_{\rm f}\tau}}{k_{\rm f} - k_b}.
\label{eq:waitDist6}
\end{align}

\subsection{Transition rates in a tilted periodic potential}

\begin{figure*}[h]
\centering
\includegraphics[width=0.3\textwidth]{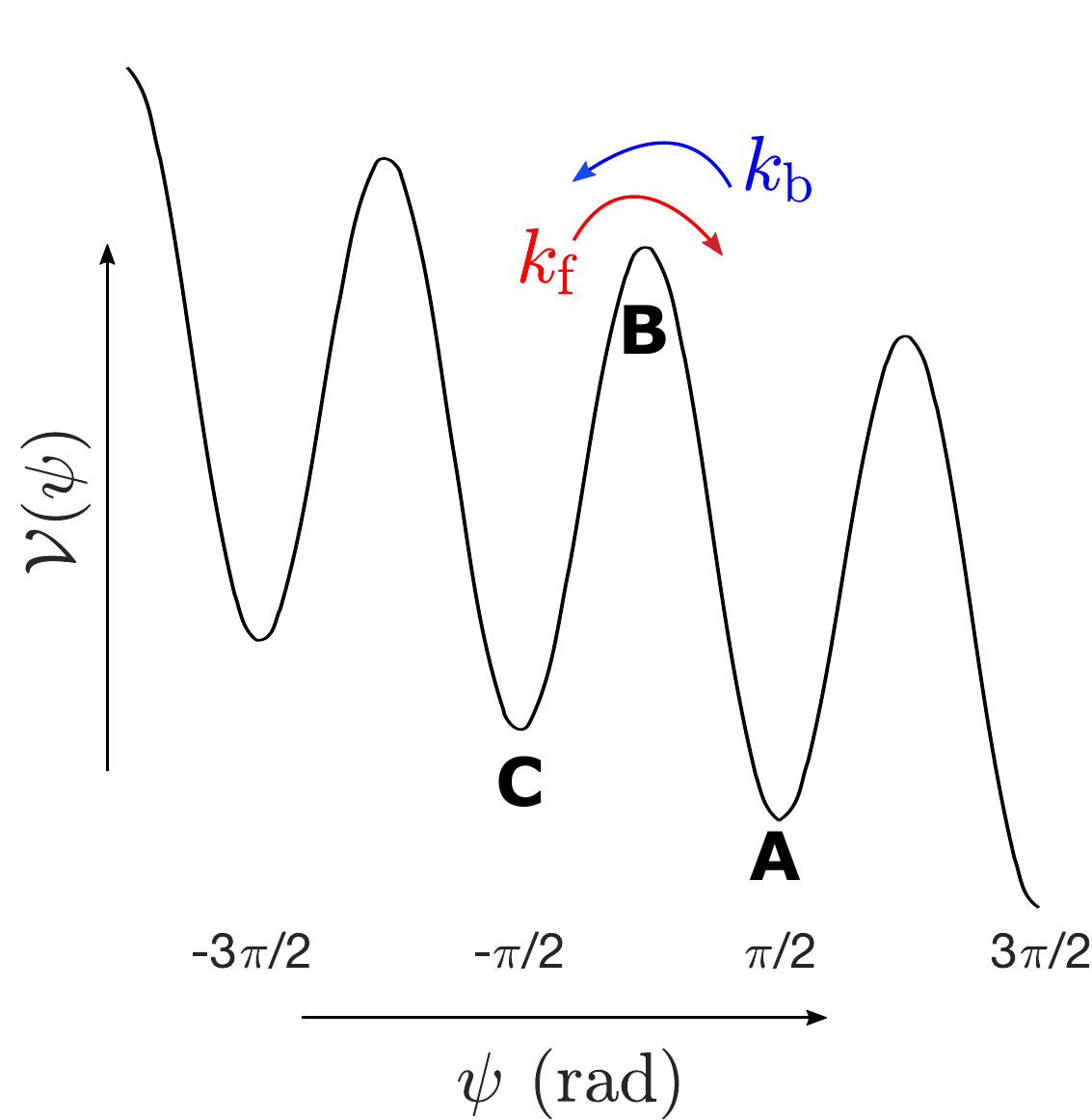}
\caption{The tilted potential used for calculating the transition rates $k_{\rm f/b}$ in the Kramers' approximation.}
\label{fig:potential}
\end{figure*}

We observe from FIG.~\ref{fig:Bayesian_Parameters}, that the active speed $v_0 \sim [4 - 8]\, mm/s$. Given that the mean-radius of the confining channel $R=116.5 \, mm$, the parameter $\nu = v_0/R$ is quite small, say compared to $\omega$. Further, we find that $\nu$ is small compared to $\lambda$ as well FIG.~\ref{fig:SI_lambda}. As such, we approximate the effective potential to
\begin{align}
\mathcal{V}(\psi)=-\omega\psi + \frac{\lambda}{2}\cos 2\psi.
\end{align}
We calculate the forward/backward (along/opposite-to $\omega$) transition rates in the Kramers' approximation for a particle moving in the potential $\mathcal{V}(\psi)$ \cite{Hanggi1990}. We calculate the potential height difference $\Delta \mathcal{V}_{BA}$ between points  A and B, and $\Delta \mathcal{V}_{BC}$ between points B and C as shown in FIG.~\ref{fig:potential}\,.
\begin{align}
\Delta \mathcal{V}_{BA} = |\mathcal{V}_B  - \mathcal{V}_A|=\sqrt{\lambda^2-\omega^2}-\omega \sin^{-1}\frac{\omega}{\lambda}+\omega\frac{\pi}{2}\,,
\end{align}
\begin{align}
\Delta \mathcal{V}_{BC} = |\mathcal{V}_B - \mathcal{V}_C|=\sqrt{\lambda^2-\omega^2}-\omega \sin^{-1}\frac{\omega}{\lambda} - \omega\frac{\pi}{2}\,.
\end{align}
The second derivative of the potential at the required points are 
\begin{align}
\mathcal{V}''(A)=2\sqrt{\lambda^2-\omega^2}=\mathcal{V}''(C)=-\mathcal{V}''(B)\,.
\end{align}
Using Kramers' escape rate theory, we then obtain the forward and backward flipping rates $k_{\rm f}$ and $k_{\rm b}$ respectively    
\begin{align}
k_{\rm f} &= \frac{\sqrt{\mathcal{V}''(C) \; |\mathcal{V}''(B)|}}{2\pi} e^{-\Delta \mathcal{V}_{BC}/D}\,,
\nonumber \\
&=\frac{\sqrt{\lambda^2-\omega^2}}{\pi} \exp\left( -\frac{\sqrt{\lambda^2-\omega^2}\,-\,\omega \sin^{-1}{(\omega/\lambda)}\,-\, \pi\omega/2}{D} \right),
\end{align}
and
\begin{align}
k_{\rm b} &= \frac{\sqrt{\mathcal{V}''(A) \; |\mathcal{V}''(B)|}}{2\pi} e^{-\Delta \mathcal{V}_{BA}/D}\,,
\nonumber \\
&=\frac{\sqrt{\lambda^2-\omega^2}}{\pi} \exp\left( -\frac{\sqrt{\lambda^2-\omega^2}\,-\,\omega \sin^{-1}{(\omega/\lambda)}\, + \, \pi\omega/2}{D} \right).
\end{align}
Thus, the ratio of the two rates
\begin{align}
\frac{k_{\rm f}}{k_{\rm b}} = e^{\pi\omega/D}
\end{align}
is controlled by the constant ``force'' $\omega$ (in our case the chirality of the active particle).

\newpage 

\section{Movies}

\textbf{Supplementary Movie 1 (00Unconfined.avi):}

\begin{figure*}[h!]
\centering
\includegraphics[width=0.5\textwidth]{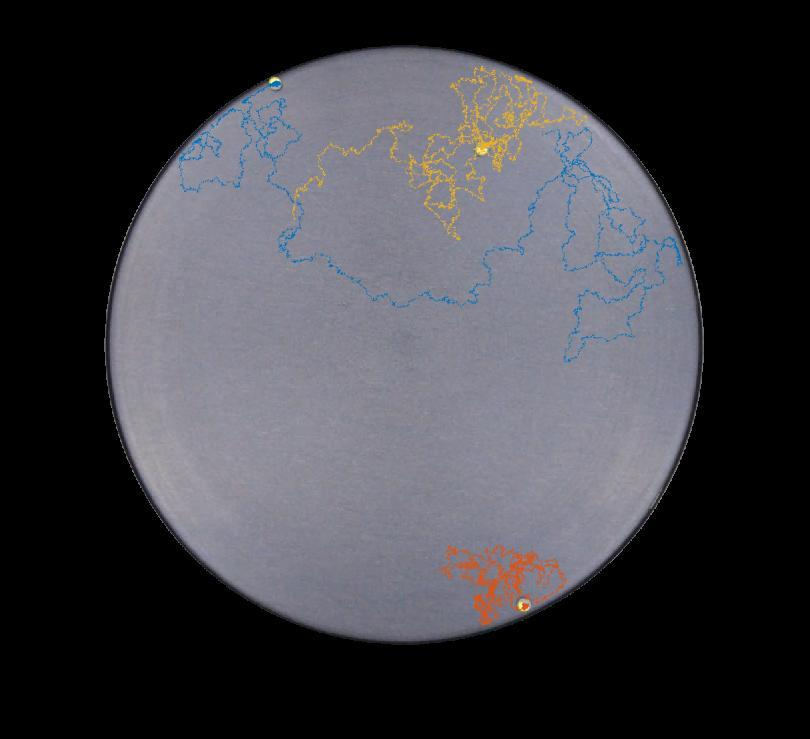}
\end{figure*}

Active self-propelled motion of disks in a 2-D arena. The particles have a diameter of 4.5~mm. Here the driving $\Gamma = 6.71$. Previous positions are continuosly overlaid to indicate the overall trajectory of the particles.

\newpage

\textbf{Supplementary Movie 2 (01ConfinedABP.avi): }

\begin{figure*}[h!]
\centering
\includegraphics[width=0.5\textwidth]{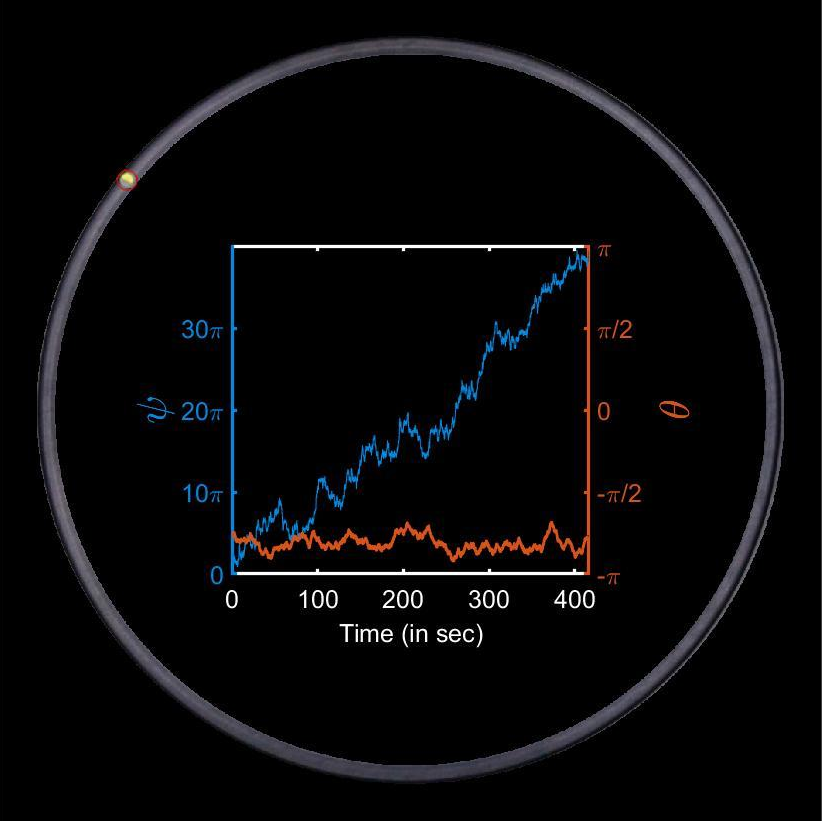}
\end{figure*}

ABP-like motion in a confined channel. Here, the particle diameter is 4.5~mm, the channel confinement $\delta = \rm 0.88$ and the driving $\Gamma =  6.71$. A red circle is used to highlight the particle. Insets show the time evolution of the position $\theta$ and the relative internal orientation $\psi$.

\newpage

\textbf{Supplementary Movie 3 (02ConfinedRTP.avi): }

\begin{figure*}[h!]
\centering
\includegraphics[width=0.5\textwidth]{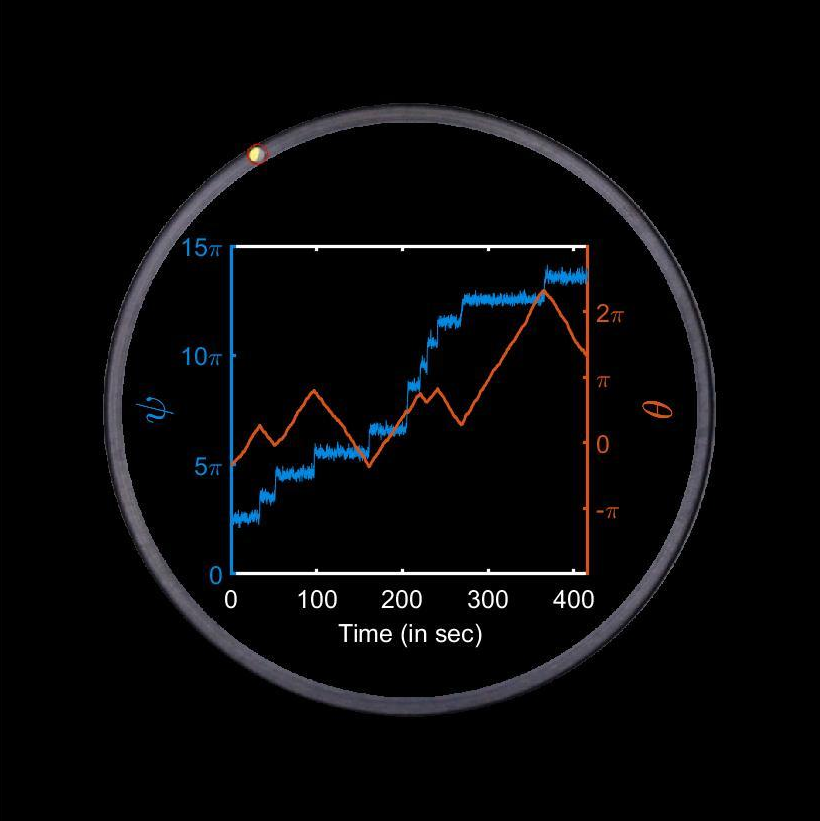}
\end{figure*}

RTP-like motion in a confined channel. Here, the particle diameter is 4.5~mm, the channel confinement $\delta = \rm 0.9$ and the driving $\Gamma = 6.71$. A red circle is used to highlight the particle. Insets show the time evolution of the position $\theta$ and the relative internal orientation $\psi$.

\end{document}